\documentclass[english, journal, 12pt, onecolumn, draftclsnofoot]{IEEEtran}

\usepackage{enumitem}

\usepackage{hyperref}
\usepackage[T1]{fontenc}
\usepackage[utf8]{inputenc}
\usepackage{verbatim}
\usepackage{float}
\usepackage[cmex10]{amsmath}
\usepackage{amssymb}
\usepackage{graphicx}
\usepackage{epstopdf}
\usepackage{psfrag}
\usepackage{stmaryrd}
\usepackage{mathrsfs}
\usepackage{cite}
\makeatletter
\floatstyle{ruled}

\@ifundefined{showcaptionsetup}{}{%
 \PassOptionsToPackage{caption=false}{subfig}}
\usepackage{subfig}
\makeatother

\newcommand{\cip}{\overset{\mathcal{P}}{\longrightarrow}}
\providecommand{\eref}[1]{\eqref{eq:#1}}  
\providecommand{\sref}[1]{Section~\ref{sec:#1}}
\providecommand{\fref}[1]{Figure~\ref{fig:#1}}

\DeclareMathOperator{\polylog}{polylog}
\def\Pc{P_\text{correct}}
\def\opts{\widetilde{\vx}}

\usepackage{babel}
\usepackage{dsfont}
\newtheorem{thm}{Theorem}
\newtheorem{prop}{Proposition}
\newtheorem{lem}{Lemma}

\newtheorem{rem}{Remark}

\begin{document}
\global\long\def\vct#1{\boldsymbol{#1}}%
\global\long\def\mat#1{\boldsymbol{#1}}%
\global\long\def\opvec{\text{vec}}%
\global\long\def\optr{\mbox{tr}}%
\global\long\def\opmat{\mbox{mat}}%
\global\long\def\opdiag{\mbox{diag}}%

\global\long\def\t#1{\widetilde{#1}}%
\global\long\def\h#1{\widehat{#1}}%
\global\long\def\abs#1{\left\lvert #1\right\rvert }%
\global\long\def\norm#1{\lVert#1\rVert}%

\global\long\def\inprod#1{\langle#1\rangle}%
\global\long\def\set#1{\left\{  #1\right\}  }%
\global\long\def\bydef{\overset{\text{def}}{=}}%
\global\long\def\teq#1{\overset{#1}{=}}%
\global\long\def\tleq#1{\overset{#1}{\leq}}%
\global\long\def\tgeq#1{\overset{#1}{\geq}}%
\global\long\def\tapprox#1{\overset{#1}{\approx}}%

\global\long\def\EE{\mathbb{E}\,}%
\global\long\def\EEk{\mathbb{E}_{k}\,}%

\global\long\def\R{\mathbb{R}}%
\global\long\def\E{\mathbb{E}}%
\global\long\def\P{\mathbb{P}}%
\global\long\def\Q{\mathbb{Q}}%
\global\long\def\I{\mathds{1}}%

\global\long\def\va{\boldsymbol{a}}%
\global\long\def\vb{\boldsymbol{b}}%
\global\long\def\vc{\boldsymbol{c}}%
\global\long\def\vd{\boldsymbol{d}}%
\global\long\def\ve{\boldsymbol{e}}%
\global\long\def\vf{\boldsymbol{f}}%
\global\long\def\vg{\boldsymbol{g}}%

\global\long\def\vh{\boldsymbol{h}}%
\global\long\def\vi{\boldsymbol{i}}%
\global\long\def\vj{\boldsymbol{j}}%
\global\long\def\vk{\boldsymbol{k}}%
\global\long\def\vl{\boldsymbol{l}}%

\global\long\def\vm{\boldsymbol{m}}%
\global\long\def\vn{\boldsymbol{n}}%
\global\long\def\vo{\boldsymbol{o}}%
\global\long\def\vp{\boldsymbol{p}}%
\global\long\def\vq{\boldsymbol{q}}%
\global\long\def\vr{\boldsymbol{r}}%

\global\long\def\vs{\boldsymbol{s}}%
\global\long\def\vt{\boldsymbol{t}}%
\global\long\def\vu{\boldsymbol{u}}%
\global\long\def\vv{\boldsymbol{v}}%
\global\long\def\vw{\boldsymbol{w}}%
\global\long\def\vx{\boldsymbol{x}}%
\global\long\def\vy{\boldsymbol{y}}%
\global\long\def\vz{\boldsymbol{z}}%

\global\long\def\vtheta{\boldsymbol{\theta}}%
\global\long\def\vxi{\boldsymbol{\xi}}%
\global\long\def\vdelta{\boldsymbol{\delta}}%
\global\long\def\veta{\vct{\eta}}%
\global\long\def\vlambda{\boldsymbol{\lambda}}%
\global\long\def\vbeta{\boldsymbol{\beta}}%
\global\long\def\vmu{\boldsymbol{\mu}}%
\global\long\def\vnu{\boldsymbol{\nu}}%

\global\long\def\mA{\boldsymbol{A}}%
\global\long\def\mB{\boldsymbol{B}}%
\global\long\def\mC{\boldsymbol{C}}%
\global\long\def\mD{\boldsymbol{D}}%
\global\long\def\mE{\boldsymbol{E}}%
\global\long\def\mF{\boldsymbol{F}}%
\global\long\def\mG{\boldsymbol{G}}%

\global\long\def\mH{\boldsymbol{H}}%
\global\long\def\mI{\boldsymbol{I}}%
\global\long\def\mJ{\boldsymbol{J}}%
\global\long\def\mK{\boldsymbol{K}}%
\global\long\def\mL{\boldsymbol{L}}%
\global\long\def\mM{\boldsymbol{M}}%
\global\long\def\mN{\boldsymbol{N}}%

\global\long\def\mO{\boldsymbol{O}}%
\global\long\def\mP{\boldsymbol{P}}%
\global\long\def\mQ{\boldsymbol{Q}}%
\global\long\def\mR{\boldsymbol{R}}%
\global\long\def\mS{\boldsymbol{S}}%
\global\long\def\mT{\boldsymbol{T}}%
\global\long\def\mU{\boldsymbol{U}}%

\global\long\def\mV{\boldsymbol{V}}%
\global\long\def\mW{\boldsymbol{W}}%
\global\long\def\mX{\boldsymbol{X}}%
\global\long\def\mY{\boldsymbol{Y}}%
\global\long\def\mZ{\boldsymbol{Z}}%

\global\long\def\mLa{\boldsymbol{\Lambda}}%
\global\long\def\mOm{\boldsymbol{\Omega}}%

\global\long\def\calS{\mathcal{S}}%
\global\long\def\calN{\mathcal{N}}%
\global\long\def\calL{\mathcal{L}}%
\global\long\def\calD{\mathcal{D}}%
\global\long\def\calV{\mathcal{V}}%
\global\long\def\calW{\mathcal{W}}%
\global\long\def\calO{\mathcal{O}}%

\global\long\def\a{\alpha}%
\global\long\def\b{\beta}%
\global\long\def\m{\mu}%
\global\long\def\n{\nu}%
\global\long\def\g{\gamma}%
\global\long\def\s{\sigma}%
\global\long\def\e{\epsilon}%
\global\long\def\w{\omega}%
\global\long\def\veps{\varepsilon}%

\global\long\def\T{\intercal}%
\global\long\def\d{\text{d}}%

\global\long\def\nt{\left\lfloor nt\right\rfloor }%
\global\long\def\ns{\left\lfloor ns\right\rfloor }%
\global\long\def\textif{\text{if }}%
\global\long\def\otherwise{\text{otherwise}}%
\global\long\def\st{\text{s.t. }}%

\global\long\def\tvy{\t{\boldsymbol{y}}}%
\global\long\def\tc{\t c}%
\global\long\def\ttau{\t{\tau}}%
\global\long\def\tf{\t f}%
\global\long\def\th{\t h}%
\global\long\def\tq{\t q}%
\global\long\def\tz{\t z}%

\global\long\def\tva{\t{\boldsymbol{a}}}%
\global\long\def\tvw{\t{\boldsymbol{w}}}%
\global\long\def\tw{\t w}%

\global\long\def\asconv{\overset{a.s.}{\rightarrow}}%
\global\long\def\pconv{\overset{\mathbb{P}}{\rightarrow}}%
\global\long\def\iid{\overset{i.i.d.}{\sim}}%
\global\long\def\dmtx{\mA}%
\global\long\def\dmtxi{A}%

\global\long\def\sgl{\boldsymbol{\beta}}%
\global\long\def\noise{\boldsymbol{w}}%
\global\long\def\sgli{\beta}%
\global\long\def\est{\vx}%
\global\long\def\esti{x}%
\global\long\def\res{\vr}%
\global\long\def\err{\vv}%

\global\long\def\noisei{w}%
\global\long\def\sol{\widehat{\sgl}}%
\global\long\def\soli{\widehat{\beta}}%
\global\long\def\equinoise{\mZ}%
\global\long\def\equinoisei{Z}%

\global\long\def\argmin#1{\underset{#1}{\text{\ensuremath{\arg\min} }}}%
\global\long\def\argmax#1{\underset{#1}{\text{\ensuremath{\arg\max}}}}%

\global\long\def\regu{J_{\vlambda}}%
\global\long\def\sregu{J}%
\global\long\def\comset{\mathcal{S}}%

\global\long\def\prox{\eta}%
\global\long\def\vprox{\veta}%
\global\long\def\tprox{\text{Prox}}%
\global\long\def\sgn{\text{sign}}%
\global\long\def\tproxl{\text{Prox}_{\vlambda}}%
\global\long\def\tproxq{\text{Prox}_{\ell_{2}}}%
\global\long\def\tproxg{\text{Prox}}%

\global\long\def\snr{\text{SNR }}%
\global\long\def\nsdp{\sigma_{p}}%
\global\long\def\nvarp{\sigma_{p}^{2}}%

\global\long\def\averG{\overline{G}}%
\global\long\def\averg{\bar{g}}%
\global\long\def\errx{v_{x}}%
\global\long\def\optx{\boldsymbol{x}^{*}}%
\global\long\def\optxi#1{x_{#1}^{*}}%
\global\long\def\optu{\boldsymbol{u}^{*}}%
\global\long\def\optui#1{u_{#1}^{*}}%

\global\long\def\lossprim{Q}%
\global\long\def\subd{v}%
\global\long\def\poeps#1{C_{#1}}%
\global\long\def\averpoeps#1{\overline{C}_{#1}}%
\global\long\def\optxsub#1{\boldsymbol{x}_{#1}^{*}}%
\global\long\def\field{\theta}%

\global\long\def\losspo#1{Q_{#1}}%
\global\long\def\averlosspo#1{\overline{Q}_{#1}}%
\global\long\def\neb{N_{e}}%
\global\long\def\tneb{\widetilde{N}_{e}}%
\global\long\def\tnew{\text{new}}%
\global\long\def\expnt{\gamma}%

\global\long\def\tauAO{\tau_{\text{AO}}^{*}}%
\global\long\def\opttau{\tau_{p}}%
\global\long\def\optS{f_{p}}%
\global\long\def\optQ{Q_{p}^{*}}%
\global\long\def\optlam{\lambda_{p}}
\global\long\def\ratiob{\delta_{L}}%
\global\long\def\sampratio{\delta_{p}}%
\global\long\def\ptline{\alpha_{p}}%
\global\long\def\limptline{\alpha^{*}}%


\global\long\def\dmtxcav{\mA}%
\global\long\def\vxcav{\vx}%
\global\long\def\vycav{\vy}%
\global\long\def\vucav{\vu}%
\global\long\def\sglcav{\sgl}%
\global\long\def\sglicav{\sgli}%
\global\long\def\xcav{x}%
\global\long\def\optxcav{\xcav^{*}}%
\global\long\def\vcav{v}%
\global\long\def\vacav{\va}%
\global\long\def\tvacav{\widetilde{\va}}%
\global\long\def\Lcav{L}%
\global\long\def\xhatcav{\breve{x}}%
\global\long\def\xtdcav{\widetilde{x}}%

\global\long\def\xhat{\hat{x}^{*}}%
\global\long\def\xtd{\tilde{x}^{*}}%
\global\long\def\optxtd{\widetilde{\vx}^{*}}%

\global\long\def\var{\text{Var}}%
\title{The Limiting Poisson Law of Massive MIMO Detection with Box Relaxation}

 \author{%
   \IEEEauthorblockN{Hong Hu and Yue M. Lu}
    \thanks{H. Hu and Y. M. Lu are with the John A. Paulson School of Engineering and Applied Sciences, Harvard University, Cambridge, MA 02138, USA (e-mails: honghu@g.harvard.edu and yuelu@seas.harvard.edu). This work was supported by the Harvard FAS Dean's Fund for Promising Scholarship, and by the US National Science Foundation under grants CCF-1718698 and CCF-1910410.}
 }

\maketitle


\begin{abstract}
Estimating a binary vector from noisy linear measurements is a prototypical problem for MIMO systems. A popular algorithm, called the box-relaxation decoder, estimates the target signal by solving a least squares problem with convex constraints. This paper shows that the performance of the algorithm, measured by the number of incorrectly-decoded bits, has a limiting Poisson law. This occurs when the sampling ratio and noise variance, two key parameters of the problem, follow certain scalings as the system dimension grows. Moreover, at a well-defined threshold, the probability of perfect recovery is shown to undergo a phase transition that can be characterized by the Gumbel distribution. Numerical simulations corroborate these theoretical predictions, showing that they match the actual performance of the algorithm even in moderate system dimensions.
\end{abstract}

\section{Introduction}

\subsection{Motivations}
Consider the problem of estimating a binary vector $\sgl\in\left\{ -1,1\right\} ^{p}$ from noisy linear measurements in the form of
\begin{equation}
\vy=\dmtx\sgl+\noise.\label{eq:linear_model}
\end{equation}
Here, $\dmtx\in\R^{n\times p}$ is a known sensing matrix 
and $\noise \sim\mathcal{N}(\boldsymbol{0}, \sigma_p^2 \mI_n)$ denotes an unknown noise vector. This is a prototypical model for multi-user detections in MIMO communication systems \cite{viterbi1995cdma, rusek2012scaling}. It also arises in other applications such as compressed sensing \cite{das2013finite}, source separation \cite{aissa2015sparsity}, and image processing \cite{ahn2016compressive}.

Various algorithms have been proposed to solve (\ref{eq:linear_model}). Examples include sphere decoding \cite{pohst1981computation}, zero-forcing \cite{grotschel2012geometric}, approximate message passing \cite{jeon2015optimality}, Markov chain Monte Carlo methods \cite{hassibi2014optimized}, and semidefinite programming \cite{luo2010semidefinite }. Among them, a convex-optimization based method, known as the box-relaxation decoder \cite{tan2000box,thrampoulidis2016ber,thrampoulidis2018symbol}, is popular in practice due to its simplicity and efficiency.
The method consists of merely two
steps: (1) solve a box-constrained least squares problem
\begin{equation}
\optx=\argmin{\vx\in[-1,1]^{p}}\frac{1}{2}\|\vy-\mA\vx\|^{2},\label{eq:box_LASSO}
\end{equation}
and (2) obtain an estimate of $\sgl$ by taking the sign of $\optx$, \emph{i.e.}, $\sol=\sgn(\optx).$ 

The performance of this algorithm can be measured by the bit error rate (BER):
\begin{equation}\label{eq:BER}
\text{BER} = \frac{1}{p}\sum_{i=1}^{p}\I_{\{\soli_{i}\neq\beta_{i}\}},
\end{equation}
where $\I_{\{\cdot\}}$ denotes the indicator function. The achievable BER depends on two key parameters: the noise variance $\sigma_p^2$, and the sampling ratio $\delta_p \bydef n/p$.

Under the assumption that the sensing matrix $\mA$ has i.i.d. normal entries, the authors of  \cite{thrampoulidis2016ber,thrampoulidis2018symbol} analyzed the asymptotic BER achieved by the box-relaxation decoder. They show that, as $n,p\to\infty$ with $\delta_p \to \delta \in (\frac{1}{2}, \infty)$ and $\nsdp^2 \equiv \sigma^2>0$, the BER converges in probability to a deterministic limit, \emph{i.e.},
\begin{equation}\label{eq:asymp_BER}
\text{BER} \cip \mathcal{E}(\delta, \sigma^2) \in\left(0, \tfrac{1}{2}\right).
\end{equation}
This means that for any $\sigma^2>0$ and $\delta>\frac{1}{2}$, the algorithm can asymptotically achieve a \emph{weak recovery} of $\sgl$: it is better than random guess, but $\sol$ always contains a nonzero fraction of errors. Moreover, one can show that 
\begin{equation}\label{eq:asymp_E}
\lim_{\delta \to \infty} \mathcal{E}(\delta, \sigma^2)  = \lim_{\sigma^2 \to 0} \mathcal{E}(\delta, \sigma^2) = 0.
\end{equation}

The expressions in \eref{asymp_E}, together with \eref{asymp_BER}, suggest that the asymptotic BER can be made arbitrarily small if we increase the number of measurements or reduce the noise variance. This then raises a tantalizing question: is there a regime of $(\delta_p, \sigma_p^2)$ such that the box-relaxation decoder can \emph{perfectly} recover the target signal? Existing results in \cite{thrampoulidis2016ber,thrampoulidis2018symbol} cannot answer this question, for two reasons. First, $\text{BER} \overset{p \to \infty}{\longrightarrow} 0$ only guarantees that the \emph{number of error bits}
\begin{equation}
\neb\bydef\sum_{i=1}^{p}\I_{\{\soli_{i}\neq\beta_{i}\}},\label{eq:NEB}
\end{equation}
is sublinear in $p$, but it contains no information about the actual distribution of $N_e$, including whether $N_e = 0$. The second issue is subtle but important. It has to do with the specific order with which the limits are taken in \eref{asymp_BER} and \eref{asymp_E}. There, we first send the dimension $p \to \infty$ \emph{before} letting $\delta_p \to \infty$ or $\sigma_p^2 \to 0$. In practice, $p$ is large but always finite, and thus the speed with which $\delta_p \to \infty$ and $\sigma_p^2 \to 0$ [\emph{e.g.}, $\sigma_p^2 = \mathcal{O}(1/p)$ vs. $\sigma_p^2 = \mathcal{O}(1/\log p)$] makes all the difference.

The goal of this paper is to present a precise asymptotic characterization of the probability distribution of $N_e$. We show that, in certain scaling regimes of $(\delta_p, \sigma_p^2)$, the distribution of $N_e$ converges to a Poisson law. Moreover, we derive conditions under which the exact recovery of $\sgl$ is possible and provide an asymptotic formula for $\P(\neb=0)$ in the form of a Gumbel distribution.

\subsection{Main Results\label{sec:Main-Results}}

We make the following assumptions throughout the paper.
\begin{enumerate}[label={(A.\arabic*)}]
\item \label{asmp:A1}The elements of $\dmtx$ are drawn from the i.i.d. Gaussian distribution: $\dmtxi_{ij} \iid \mathcal{N}(0,\,\frac{1}{p})$.
\item \label{asmp:Abeta} $\sgl = - \boldsymbol{1}_p$, where $\boldsymbol{1}_p$ denotes the all-ones vector.
\item \label{asmp:A2}The noise is Gaussian: $\noise \sim \mathcal{N}(\boldsymbol{0},\nvarp \mI_n)$.
\item \label{asmp:A3}$\liminf_{p\to\infty}\sampratio > 1/2$ and $\limsup_{p\to\infty}\sampratio/\log p < \infty$.
\item \label{asmp:A4}$\liminf_{p\to\infty}\nsdp^2 \log^2 p > 0$ and $\limsup_{p\to\infty}\nsdp^2 < \infty$.
\end{enumerate}

In \ref{asmp:Abeta}, we assume that each coordinate of true signal is $-1$ to simplify our derivations. All the results still hold for arbitrary $\sgl$, due to the rotational symmetry of $\dmtx$.  In \ref{asmp:A3}, the requirement that $\liminf_{p\to\infty}\sampratio > 1/2$ is related to the fundamental limits of convex relaxation for structural signal reconstruction. In \cite{chandrasekaran2012convex}, it is shown that, if $\limsup_{p\to\infty}\sampratio\leq \frac{1}{2}$, the box-relaxation decoder cannot successfully recover $\sgl$ even in the noiseless case. In \ref{asmp:A4}, we essentially require $\nsdp^2 > c / \log^2 p$ for some $c > 0$. This restriction is due to the limitations of our current proof techniques. We expect that many of our results still hold without this restriction.


To state our main results, we first need to introduce the following potential function:
\begin{equation}\label{eq:F}
F_p(\tau; \nsdp^2, \sampratio) =\frac{\tau}{2}\left(\sampratio-\frac{1}{2}\right)+\frac{\nsdp^{2}}{2\tau}+\frac{\tau}{2}\int_{\frac{2}{\tau}}^{\infty}\left(x-\frac{2}{\tau}\right)^{2}\Phi(dx),
\end{equation}
where $\Phi$ is the CDF of the standard normal distribution. One can verify that $F_p$ is a strictly convex function of $\tau \in (0, \infty)$. (See Appendix~\ref{appen:1Dopt} for details.) Thus, one can uniquely define
\begin{equation}\label{eq:tau}
\optS \bydef \min_{\tau > 0} F_p(\tau; \nsdp^2, \sampratio) \hspace{1em} \text{and} \hspace{1em} \opttau \bydef \argmin{\tau > 0} F_p(\tau; \nsdp^2, \sampratio).
\end{equation}
Another quantity that will be crucial in our analysis is
\begin{equation}\label{eq:lambda}
\optlam \bydef p\Phi(-\tfrac{1}{\opttau}).
\end{equation}

\begin{thm}
\label{thm:PT_limitlaw_Ne}
Under \ref{asmp:A1}-\ref{asmp:A4}, and if $\limsup_{p\to\infty} \tfrac{\optlam}{\sqrt{\log p}} < \infty$, then
\begin{equation}
\label{eq:TV_inf}
d_{\text{TV}}(\neb,\mathscr{P}(\optlam)) \le \frac{\polylog p}{p^{1/5}},
\end{equation}
where $d_{\text{TV}}$ is the total variation (TV) distance and $\mathscr{P}(\lambda)$ denotes a Poisson distribution with parameter $\lambda$.
\end{thm}
\begin{rem}
The theorem, whose proof can be found in \sref{mainthmproof}, characterizes the asymptotic distribution of $\neb$ under certain scaling regimes of $(\sampratio, \nsdp^2)$. It shows that the law of $N_e$ converges to that of a Poisson random variable with parameter $\lambda_p$, if $\optlam$ grows no faster than $\sqrt{\log p}$. This requirement on $\lambda_p$ is not satisfied in the setting studied in \cite{thrampoulidis2016ber} where both $\sampratio$ and $\nsdp^2$ are kept as fixed constants and consequently $\lambda_p = \mathcal{O}(p)$. In that case, one can expect that $\sqrt{p}[\tfrac{\neb}{p}-\Phi(-\tfrac{1}{\opttau})]$ converges to a Gaussian distribution.
\end{rem}


The fact that $N_e$ can have a limiting Poisson law is not surprising. Recall from its definition in \eref{NEB} that $N_e$ is a sum of $p$ Bernoulli random variables $\{\I_{\{\soli_{i}\neq\beta_{i}\}}\}$. Moreover, one can show that $\P(\soli_{i}\neq\beta_{i}) \approx \Phi(-\tfrac{1}{\opttau})$ and that these Bernoulli random variables are close to being independent. Consequently, the law of $N_e$ is approximately a Binomial distribution $\mathcal{B}(p, \Phi(-\tfrac{1}{\opttau}))$ with an expected value equal to $\lambda_p$. As $p \to \infty$ with $\lambda_p = \mathcal{O}(\sqrt{\log p})$, it is well-known that the Binomial distribution converges to a Poisson distribution (\emph{i.e.}, the ``law of small numbers''). The technical contribution of this paper is to make the above arguments precise and rigorous. The main tool we use is the leave-one-out approach (see, \emph{e.g.}, \cite{el2018impact}), also known as the cavity method in statistical physics \cite{luo2001cavity,ramezanali2015cavity}. It allows us to carry out a detailed probabilistic analysis of the random optimization problem in \eref{box_LASSO}.
\begin{figure}
	\centering
	\begin{centering}
\psfrag{$\alpha$}{$\delta$}
\subfloat[$p = 200$\label{fig:Poisson_PMF200}]{\begin{centering}
\includegraphics[scale=0.29]{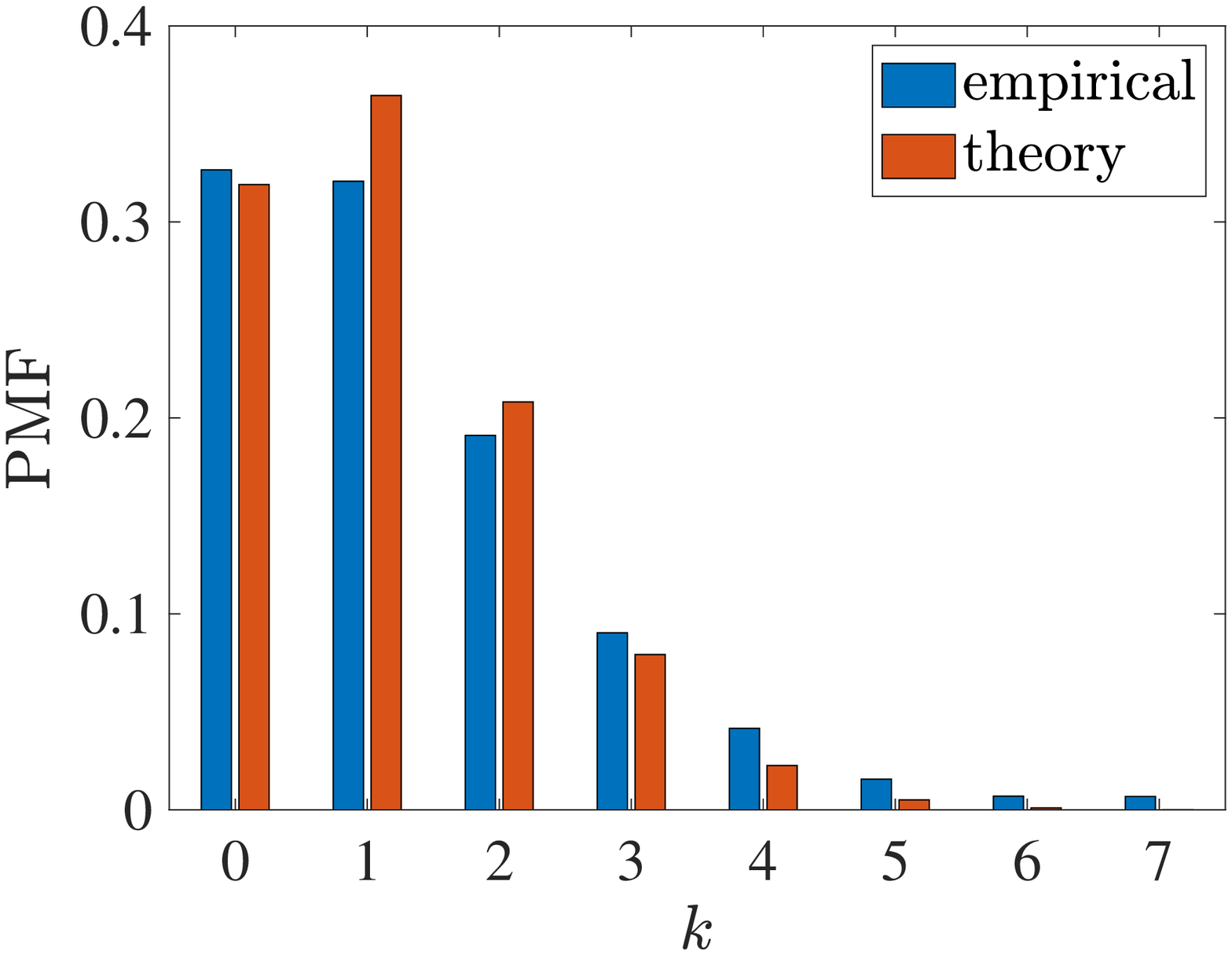}
\par\end{centering}
}\hphantom{}\subfloat[$p = 1000$\label{fig:Poisson_PMF1000}]{\begin{centering}
\includegraphics[scale=0.29]{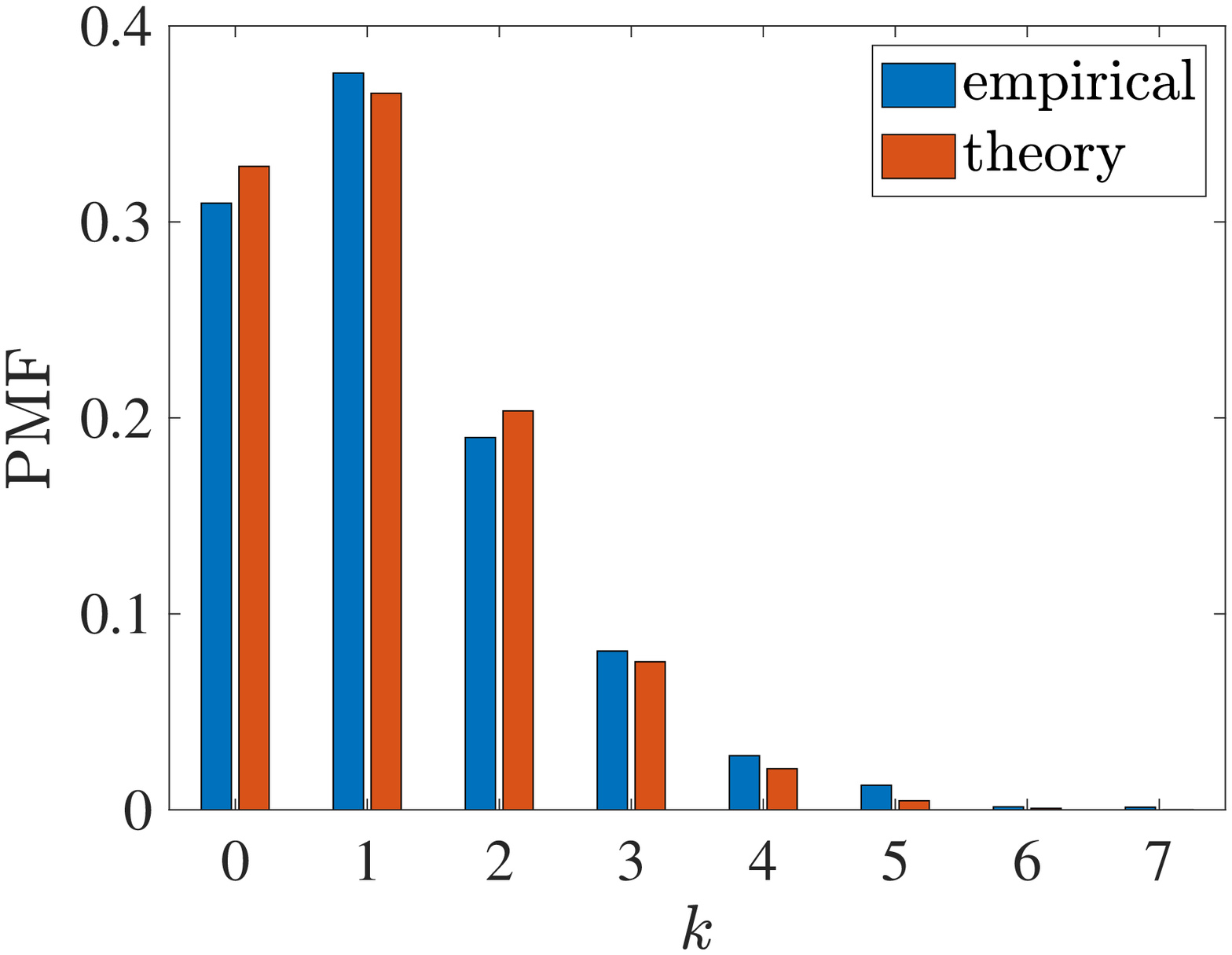}
\par\end{centering}
}\hphantom{}\subfloat[$p = 10000$\label{fig:Poisson_PMF10000}]{\begin{centering}
\includegraphics[scale=0.29]{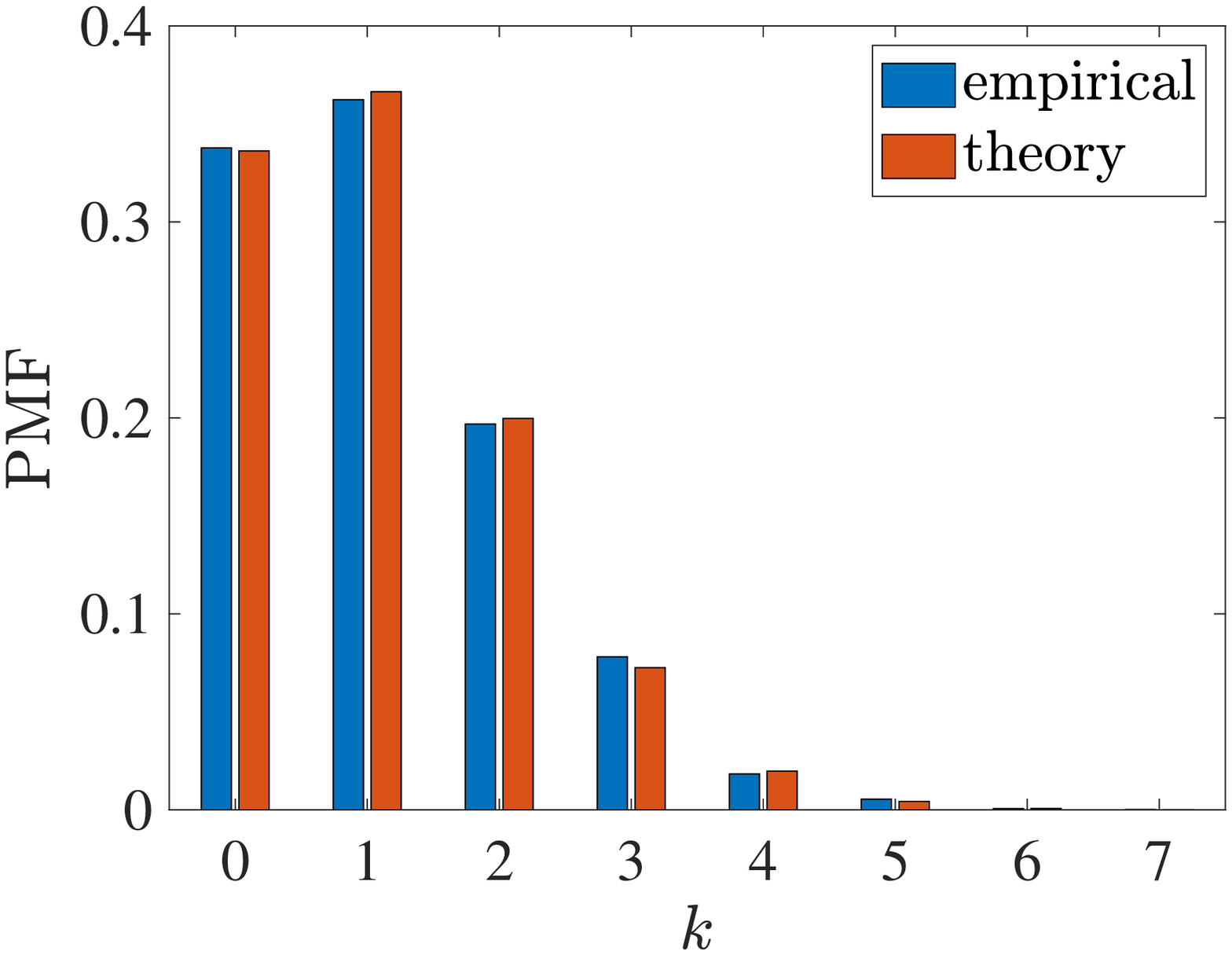}
\par\end{centering}
}
\par\end{centering}
\caption{
\label{fig:Poisson_PMF}Comparison of the empirical distribution of $N_e$ and the limiting Poisson distribution, over three different problem dimensions. In the experiments, we set $\nsdp^2 = 1$ and choose $\delta_p$ so that $\lambda_p \approx 1.1$ for all three values of $p$.
}\end{figure}

In our proof of Theorem~\ref{thm:PT_limitlaw_Ne}, we did not attempt to optimize the rate of convergence shown on the right-hand side of \eref{TV_inf}. The actual rate is likely to be faster. In \fref{Poisson_PMF}, we compare the empirical distribution of $N_e$, obtained after averaging over $10^4$ independent trials, against the limiting Poisson distribution for three different problem dimensions. We can see that, even at a moderate dimension of $p = 200$, the Poisson approximation is already accurate.

The characterization given in Theorem~\ref{thm:PT_limitlaw_Ne} allows us to study the conditions under which the box-relaxation decoder can perfectly recover the target signal. Let $\Pc \bydef \P(N_e = 0)$ denotes the probability of perfect recovery. We can show that a phase transition of $\Pc$ emerges when the following quantity
\begin{equation}\label{eq:PTline}
\ptline \bydef \frac{\sampratio-1/2}{2\nsdp^{2}\log p}
\end{equation}
is near $1$.
\begin{prop}
\label{prop:PT}
Under \ref{asmp:A1}-\ref{asmp:A4}, and if $\lim_{p\to\infty} \ptline = \limptline$, then
\begin{equation}
\lim_{p\to\infty}\Pc=\begin{cases}
1, & \text{if }\limptline > 1,\\
0, & \text{if }\limptline < 1.
\end{cases}\label{eq:Ne_transition}
\end{equation}
If $\limptline = 1$, a more refined characterization is available. Specifically, assume that
\begin{equation}\label{eq:alphap_nearPT}
\ptline(x) = 1 - \frac{\log\log p}{2\log p} + \frac{x - \log\sqrt{4\pi} }{\log p},
\end{equation}
for some constant $x\in\R$ (and thus $\ptline(x) \overset{p\to\infty}{\longrightarrow} 1$), then
\begin{equation}
\label{eq:Ne_2ndOrder_transition}
\lim_{p\to\infty}\Pc= e^{-e^{-x}},
\end{equation}
where the right-hand side is the CDF of the Gumbel distribution.
\end{prop}
\begin{rem}
The above proposition, proved in \sref{PT}, characterizes the scaling regimes of $(\sampratio, \nsdp^2)$ over which perfect recovery is achievable. The possible scalings are also flexible. For example, if we keep the sampling ratio $\sampratio$ at a fixed value $\delta > 1/2$, it then follows from \eref{PTline} and \eref{Ne_transition} that $\nsdp^2 = \frac{\delta - 1/2}{2 \log p}$ is the critical noise variance threshold for perfect recovery to happen. Alternatively, if we fix the noise variance $\nsdp^2 \equiv \sigma^2$, then the critical threshold for the sampling ratio is $\sampratio = 1/2 + 2 \sigma^2 \log p$.
\end{rem}


\begin{figure}
\begin{centering}
\psfrag{$\alpha$}{$\delta$}
\subfloat[\label{fig:phasediagram}]{\begin{centering}
\includegraphics[scale=0.38]{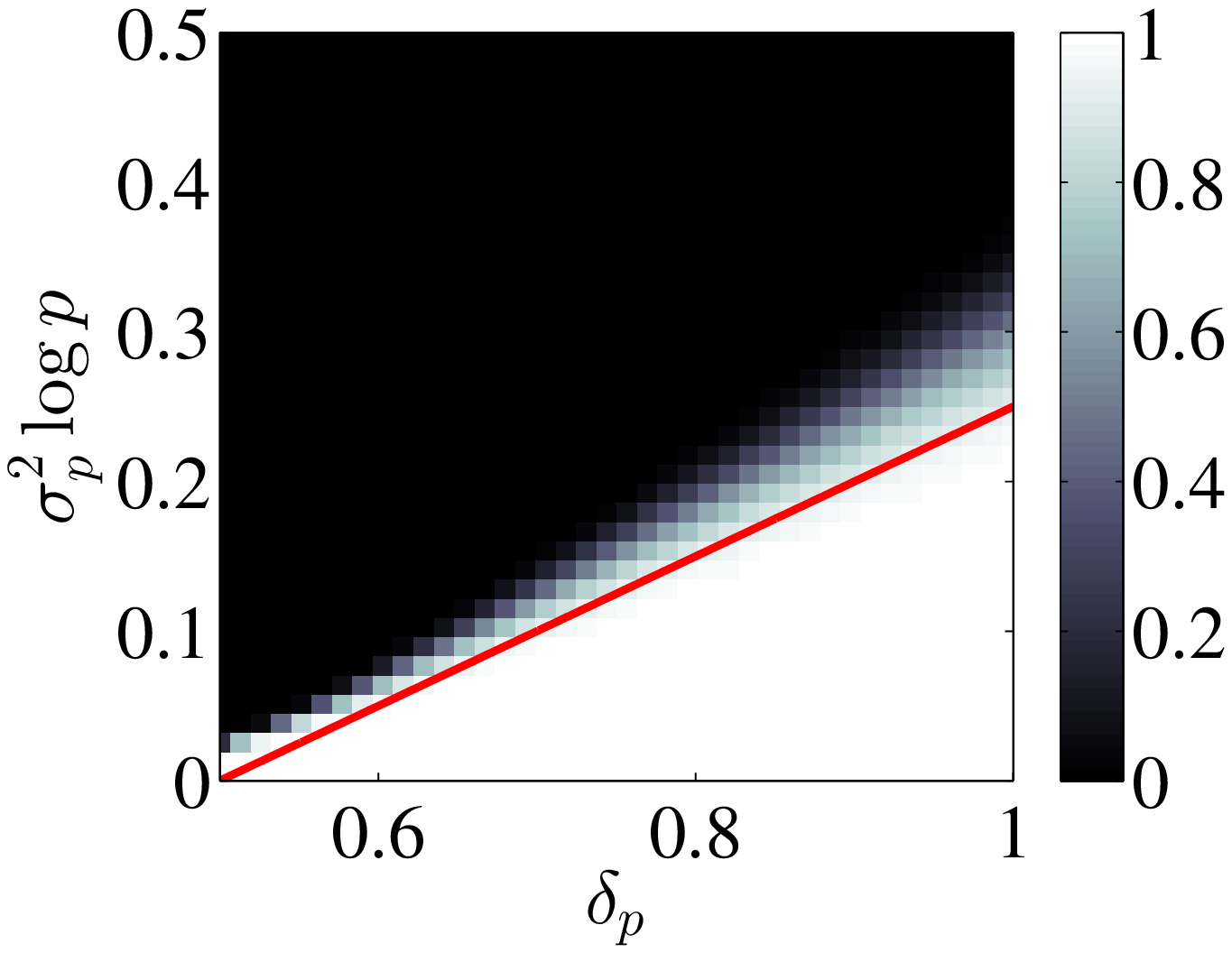}
\par\end{centering}
}\hphantom{}\subfloat[\label{fig:1stOrderTransition}]{\begin{centering}
\includegraphics[scale=0.38]{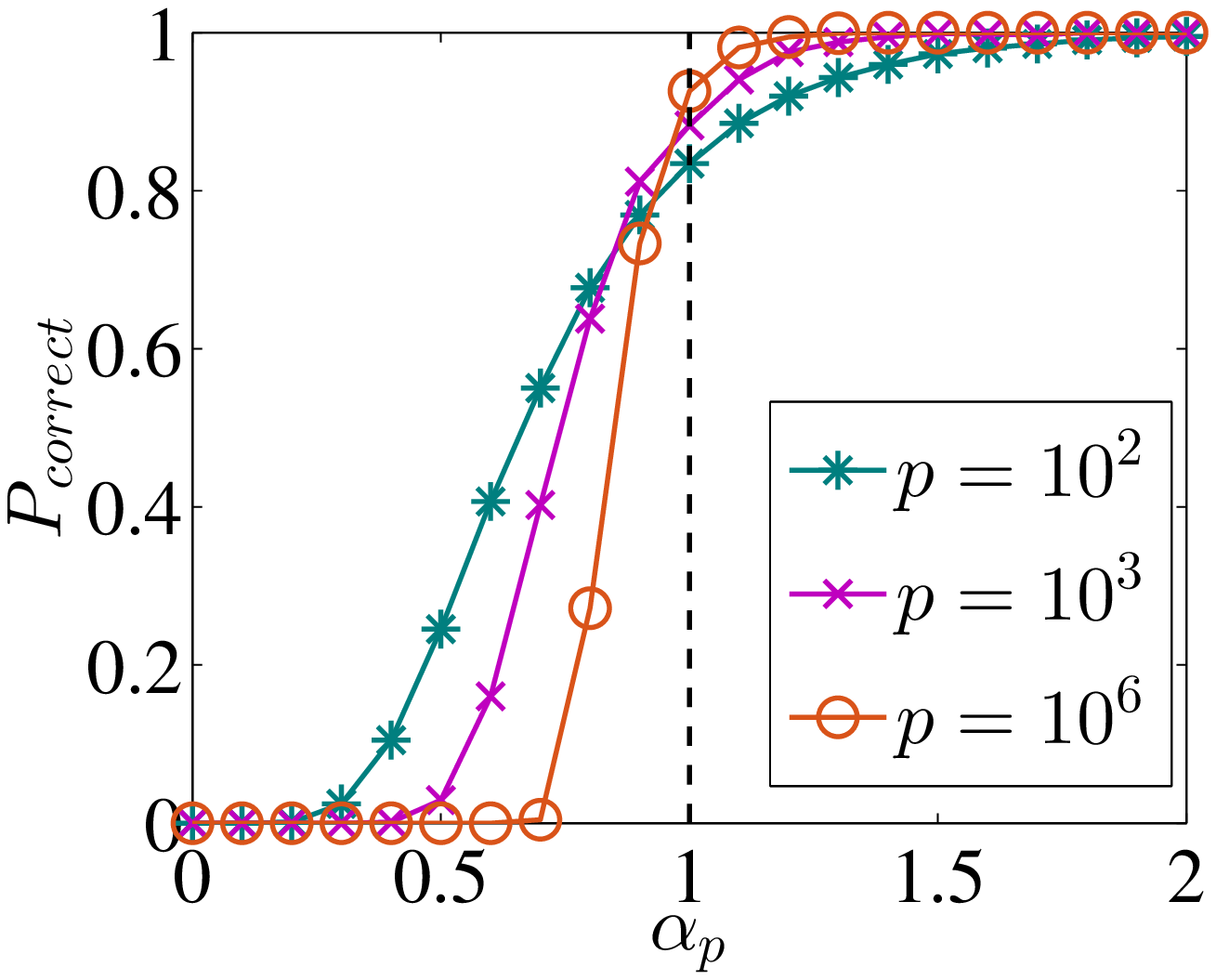}
\par\end{centering}
}\hphantom{}\subfloat[\label{fig:2ndOrderTransition}]{\begin{centering}
\includegraphics[scale=0.38]{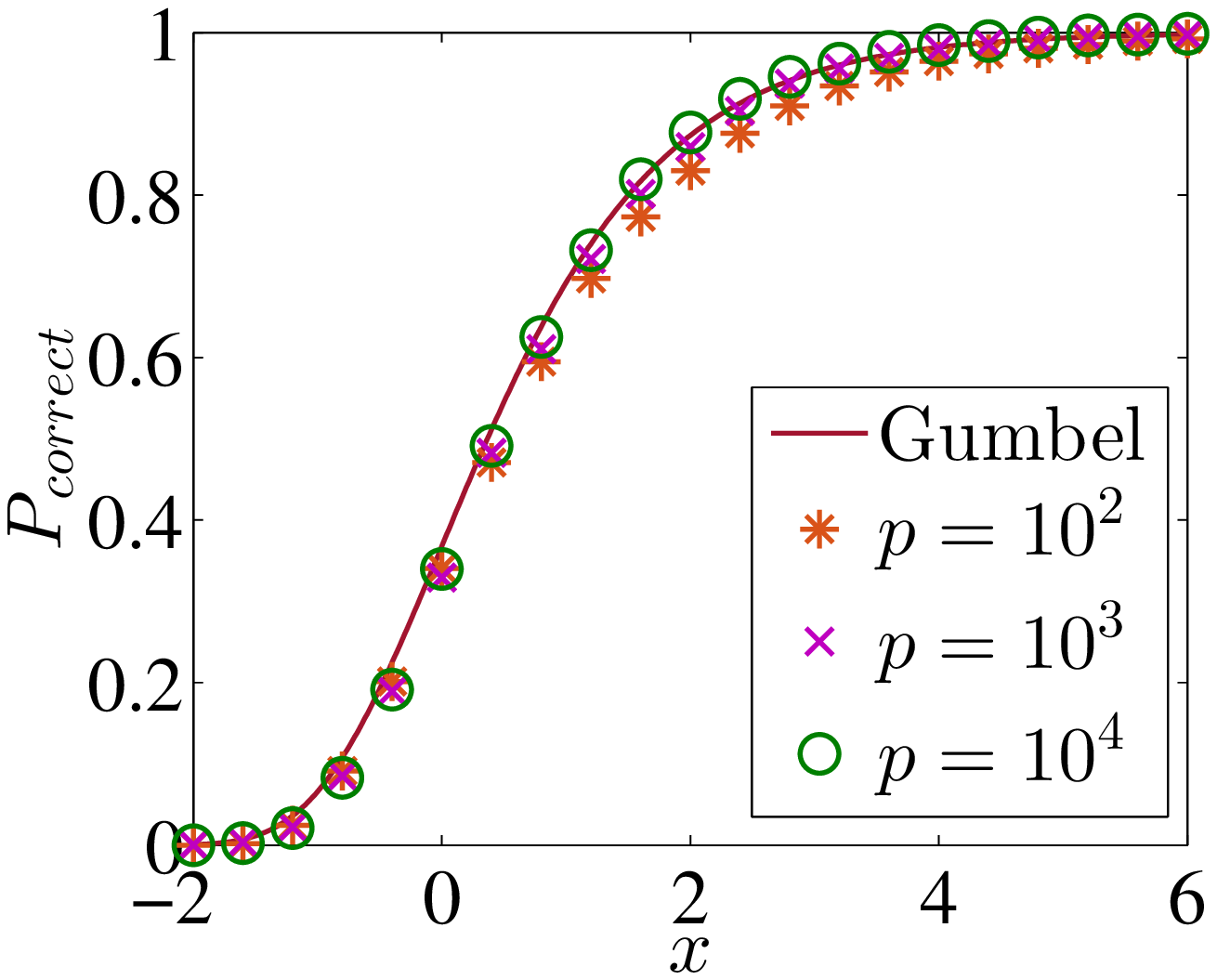}
\par\end{centering}
}
\par\end{centering}
\caption{
\label{fig:phasetransitionplot} (a) Phase diagram of the box-relaxation decoder. Each pixel represents
the value of $\Pc$ under a specific $(\sampratio,\protect\nvarp)$. The red curve is
the theoretical transition boundary: $\sigma_p^{2}\log p=\frac{\sampratio-1/2}{2}$.
(b) Phase transition of $\Pc$ with respect to $\ptline$. The dashed line represents
the theoretical threshold. (c) Near the phase transition boundary, $\Pc$ is well-approximated by the Gumbel distribution. In all three experiments, $\Pc$ is estimated by averaging over $10^4$ independent trials. In (b) and (c), we fix $\sampratio=1$ and vary $\alpha_p$ and $x$ by changing $\nsdp^2$.}
\end{figure}

To illustrate Proposition~\ref{prop:PT}, we show some results from
numerical experiments. In \fref{phasediagram}, we plot the
phase diagram of the empirical values of $\Pc$ under different choices of $(\sampratio,\nsdp^2)$, as well as the theoretical
phase transition boundary separating the regimes of perfect/nonperfect
recovery. In \fref{1stOrderTransition}, we plot $\Pc$ as a function of $\ptline$ (by fixing $\sampratio = 1$ and varying $\nsdp^2$). A transition indeed takes place near $\ptline = 1$, and the transition becomes sharper as we increase the problem dimension $p$. When $p$ is not very large, a more accurate approximation of $\Pc$ is given by the Gumbel distribution. This is illustrated in \fref{2ndOrderTransition}, where we zoom in the region near the phase
transition and compare the empirical success probability against the theoretical prediction given in \eref{Ne_2ndOrder_transition}.

%

\subsection{Related Work}

The precise analysis of high-dimensional signal estimation has already been the subject of a vast literature. Underpinning these rich results are several powerful techniques developed over the years, including the nonrigorous replica method from statistical physics \cite{tanaka2002statistical,kabashima2009typical,zdeborova2016statistical}, approximate message passing (AMP) \cite{bayati2011dynamics,donoho2009message,rangan2019vector}, the cavity method \cite{luo2001cavity,ramezanali2015cavity} and leave-one-out analysis \cite{el2018impact}, Gaussian min-max theorem (GMT) \cite{stojnic2013framework,thrampoulidis2018precise}, as well as the geometric framework based on Gaussian width \cite{chandrasekaran2012convex} and statistical dimensions \cite{amelunxen2014living}.

The box-constrained least square problem in \eref{box_LASSO} has been previously analyzed in \cite{thrampoulidis2016ber,thrampoulidis2018symbol} using GMT techniques. Analysis of similar problems can also be carried out by AMP \cite{jeon2015optimality}. However, these existing studies consider the setting where both the sampling ratio $\sampratio$ and the noise variance $\nsdp^2$ are kept as constants as $p \to \infty$. Under such scalings, one can establish that the empirical measure of $\optx$, defined as $\widehat{\mu}(\optx) \bydef \frac{1}{p}\sum_{i=1}^{p} \delta_{\optxi{i}}$, converges to some deterministic limiting measure. However, the convergence of the empirical measure is insufficient for our purpose: flipping the signs of $o(p)$ entries of $\optx$ will completely change the number of error bits $N_e$, but it has no effect on the limiting empirical measure. In view of this, we choose to use the leave-one-out approach, which allows us to construct a surrogate of $\optx$, denoted by $\opts$, in our analysis. We show that $\|\optx-\opts\|_\infty \to 0$ but the statistical properties of $\opts$ are much easier to obtain. We will elaborate on this point in Sec. \ref{sec:Proof-Sketch}.

Our work considers settings where $(\sampratio, \nsdp^2)$ can scale with the problem dimension $p$. Similar settings with flexible scalings have been explored in other contexts, including, \emph{e.g.}, sparse linear regression \cite{wainwright2009sharp,david2017high,reeves2019all}, spiked matrix estimation \cite{barbier20190}, and low-rank matrix recovery \cite{abbe2015exact}. These studies established the precise conditions under which perfect recovery in these problems is achievable. In our work, we go one step further by establishing the asymptotic distribution of the number of error bits $N_e$.

\section{Roadmap of Analysis\label{sec:Proof-Sketch}}

This section provides a general roadmap to our proof of Theorem~\ref{thm:PT_limitlaw_Ne}, which is given in \sref{mainthmproof}. To emphasize readability, we only highlight the main ideas and key intermediate results here, leaving heavier technical details to the subsequent sections and to the appendix.

\subsection{An Equivalent Scalar Problem}
\label{sec:ScalarProblemOV}
To analyze $\neb$, we need to understand the statistical properties of $\optx$, \emph{i.e.}, the optimal solution of (\ref{eq:box_LASSO}). A basic challenge lies in the fact $\optx$ is a high-dimensional vector with no closed-form expressions. The key idea behind the cavity approach \cite{luo2001cavity,ramezanali2015cavity} or the leave-one-out analysis \cite{el2018impact} is to circumvent this issue by focusing instead on a single coordinate of $\optx$. Specifically, to study the $i$th coordinate $x_i$, we can first rewrite the original problem (\ref{eq:box_LASSO}) as
\begin{align}
& \argmin{  x_i\in [-1, 1] }\;\min_{ \vx_{\backslash i} \in[-1,1]^{p-1}}\frac{1}{2}\| \dmtx_{\backslash i} \vx_{\backslash i} + \va_i(x_i-\sgli_i) - \vy_{\backslash i}\|^{2}\nonumber\\
= & \argmin{ x_i\in [-1, 1] }\;\min_{ \vx_{\backslash i} \in[-1,1]^{p-1}}\;\max_{\vu}\;\vu^{\T}[\dmtx_{\backslash i} \vx_{\backslash i} + \va_i(x_i-\sgli_i) - \vy_{\backslash i}]-\frac{1}{2}\|\vu\|^{2}\label{eq:optu_dual}\\
\teq{}  & \argmin{x_i\in[-1,1]}\max_{\vu}\;\vacav_i^{\T}\vu (x_i-\sgli_i) -L_i(\vu),\label{eq:opti1}
\end{align}
where $\vx_{\backslash i}$ is the vector formed by removing $x_i$ (and $\vbeta_{\backslash i}$ is defined in the same way),  $\va_i$ is the $i$th column of $\dmtx$, $\dmtx_{\backslash i}$ denotes the matrix formed by removing $\va_i$ from $\dmtx$, $\vy_{\backslash i} = \dmtx_{\backslash i} \sgl_{\backslash i} + \noise$,
and
\begin{equation}\label{eq:L_u}
L_i(\vu) = \|\dmtx_{\backslash i}^{\T}\vu\|_{1}+\vu^{\T}\vy_{\backslash i} + \frac{1}{2}\|\vu\|^{2}.
\end{equation}
In reaching \eref{opti1}, we have also used Sion's minimax theorem \cite{sion1958general} to swap the inner minimization and maximization in \eref{optu_dual}.

Let $\vu_{\backslash i}^{*}=\arg\,\min_{\vu}L_i(\vu)$ and define a function
\begin{equation}\label{eq:L_tilde}
g_{p,i}(v) \bydef \max_{\vu} \ (\vu - \vu_{\backslash i}^{*})^{\T} \va_i v - [{L}_i(\vu) - {L}_i(\vu_{\backslash i}^{*})].
\end{equation}
We can then check that the optimization problem (\ref{eq:opti1}) has the same solution as
\begin{align}
\argmin{x_i\in[-1,1]} g_{p,i}(x_i - \sgli_i) + \va_{i}^{\T}\vu_{\backslash i}^{*}(x_i - \sgli_i).\label{eq:scalar_problem_1a}
\end{align}
Thus, starting from the original problem \eref{box_LASSO} and after optimizing over all the ``nuisance'' variables $\vx_{\backslash i}$, we have reached in \eref{scalar_problem_1a}, an equivalent scalar optimization problem over $x_i$.

To nonspecialists, the reformulations leading to \eref{scalar_problem_1a} might look slightly mysterious, but there are several good reasons for doing so. First, note that (\ref{eq:scalar_problem_1a}) is obtained by subtracting $-{L}_i(\vu_{\backslash i}^{*})$ from (\ref{eq:opti1}). This manipulation does not change the minimizer of (\ref{eq:opti1}), but it sets the magnitude of (\ref{eq:scalar_problem_1a}) to be $\mathcal{O}(1)$, which facilitates our later analysis. Second, we explicitly pull out $\va_{i}^{\T}\vu_{\backslash i}^{*}$ in (\ref{eq:scalar_problem_1a}), since its distribution is much easier to characterize than $\va_{i}^{\T}\vu^{*}$ in (\ref{eq:opti1}), due to the independence between $\va_{i}$ and $\vu_{\backslash i}^{*}$. This is in fact a major benefit of the leave-one-out analysis. Third, as we will show next $g_{p,i}(x_i - \sgli_i)$, which is a random one-dimensional function $g_{p,i}(v)$ evaluated at $v = x_i - \sgli_i$, has a particularly simple limiting form as $p \to \infty$.

\subsection{A Limiting Quadratic Function}
\label{sec:QuadraticOV}

The following proposition, whose proof is given in \sref{scalarCP}, shows that $g_i(v)$ uniformly converges to a simple quadratic function.

\begin{prop}
\label{prop:gp_conv_2}
Under \ref{asmp:A1}-\ref{asmp:A4}, $\text{ there exists } c>0$
such that for any $i\in [p]$ and $\veps>0$,
\begin{equation}
\P\Big\{ \sup_{v\in[-2, 2]}\Big|g_{p,i}(v)-\frac{1}{2}A_{p} v^2\Big|>\varepsilon\Big\} \leq\frac{c\sampratio}{\veps}e^{-c^{-1}p\min\left\{ \tfrac{\veps^{2}}{\sampratio},\veps\right\} },\label{eq:gp_conv_2}
\end{equation}
where
\begin{equation}
A_{p}=\frac{\E\noise^{\T}(\vycav-\dmtxcav\vxcav^{*})}{\nsdp^{2}p}.\label{eq:Ap}
\end{equation}
Moreover, for $\expnt>2$ and all large enough $p$, $|A_{p}-A_{p}^{*}|<cp^{-1/(2\expnt)}$, where
\begin{equation}\label{eq:Apstar}
A_{p}^{*} \bydef {\optS}/{\opttau},
\end{equation}
and $\optS$ and $\opttau$ are the quantities defined in (\ref{eq:tau}).
\end{prop}

There is a simple intuitive explanation for why $g_{p,i}(v)$ is approximately a quadratic function. Recall that $\vu_{\backslash i}^{*}$ is the minimizer of ${L}_{i}(\vu)$. Thus, in a local neighborhood near $\vu_{\backslash i}^{*}$, we can approximate ${L}_{i}(\vu)$ by a second-order Taylor expansion: ${L}_{i}(\vu)\approx {L}_{i}(\vu_{\backslash i}^{*}) + \frac{\boldsymbol{\delta}^{\T} \mH_{\backslash i} \boldsymbol{\delta} }{2}$, where $\boldsymbol{\delta} = \vu - \vu_{\backslash i}^{*}$ and $\mH_{i}$ corresponds to the Hessian of ${L}_{i}(\vu)$ at $\vu_{\backslash i}^{*}$. Substituting this approximation into \eref{L_tilde}, we can immediately obtain that $g_{p,i}(v)\approx \frac{\va_i^{\T} \mH_{i}^{-1} \va_i}{2}v^2$.
Since $\va_i\sim \mathcal{N}(\boldsymbol{0},\tfrac{\mI_{n}}{p})$ and it is independent of $\mH_{i}$ due to the leave-one-out construction, we can expect $\va_i^{\T} \mH_{i}^{-1} \va_i$ to concentrate near a constant as $p \to \infty$. Of course, the above explanation is not rigorous in that ${L}_{i}(\vu)$ is not smooth and $\mH_{i}$ may not exist. This is one technical challenge we address in the proof.

Since $\frac{1}{2}A_{p}^{*} v^2$ is a good approximation of $g_{p,i}(v)$, we can now approximate the optimization problem in \eref{scalar_problem_1a} by
\begin{align}
\xtdcav_i &= \argmin{x_i\in[-1,1]} \frac{A_{p}^{*} (x_i - \sgli_i)^2}{2} + \va_i^{\T}\vu_{\backslash i}^{*}(x_i - \sgli_i)\nonumber \\
&= \tprox_{[-1,1]}\left(\sgli_i-\tfrac{\va_i^{\T}\vu_{\backslash i}^{*}}{A_{p}^{*}}\right), \label{eq:optxapprox2}
\end{align}
where $\tprox_{[-1,1]}$ denotes the proximal operator of the indicator function on $[-1,1]$. Its solution, denoted by $\xtdcav_i$, provides a good surrogate of $x_i^\ast$, as shown in the following proposition.

\begin{prop}
\label{prop:approximation_xi_xtildei}Under \ref{asmp:A1}-\ref{asmp:A4}, for any $\gamma>2$, there exists $c>0$, such that, for any $i\in[p]$ and $\veps\in(0,1)$,
\begin{equation}
\P\left(\left|\optxcav_i -\xtdcav_i\right|>\veps\right)<\frac{c}{\veps^2}e^{-p^{\tfrac{1}{\expnt}}\veps^{2}/c},\label{eq:conv_xistar}
\end{equation}
\end{prop}

We prove this result in \sref{convoptxCP}. Here, we demonstrate the accuracy of the approximations stated in \eref{gp_conv_2} and \eref{conv_xistar} via numerical results shown in \fref{LOO}.

Thanks to the independence between $\va_i$ and $\vu_{\backslash i}^\ast$, the surrogate solution $\xtdcav_i$ is much easier to analyze than $\optxcav_i$. Accordingly, we can consider the following approximations of $\widehat{\sgl}$ and $\neb$:
\begin{align}
\widetilde{\sgl} \bydef\sgn(\widetilde{\vx})\hspace{1em}\text{and}\hspace{1em}\tneb \bydef\sum_{i=1}^{p}\I_{\widetilde{\sgli}_{i}\neq\beta_{i}}.\label{eq:proxy_beta_hat}
\end{align}
Applying a union bound to \eref{conv_xistar} gives us $\max_i \abs{\optxi i - \xtdcav_i } \cip 0$, \emph{i.e.}, the surrogate vector $\widetilde{\vx}$ is close to ${\vx^\ast}$ in $\ell_{\infty}$ distance. This then allows us to show that $\P(\sol\neq \widetilde{\sgl})\to 0$, which also implies $d_{\text{TV}}(\neb,\tneb)\to 0$.
\begin{prop}
\label{prop:TVconvergence}
Under \ref{asmp:A1}-\ref{asmp:A4}, it holds that
\begin{equation}\label{eq:betahat_bata}
\P(\widetilde{\sgl}\neq\widehat{\sgl}) \le \lambda_{p} p^{-{1}/{5}}\polylog p,
\end{equation}
and accordingly,
\begin{equation}
d_{\text{TV}}(\neb,\tneb) \le \lambda_{p} p^{-{1}/{5}}\polylog p.\label{eq:dTV_tNe_Ne_bd_1}
\end{equation}
\end{prop}

The proof of Proposition \ref{prop:TVconvergence} can be found in \sref{convoptxCP1}. It shows that the distribution of $\neb$ is well captured by that of $\tneb$. Therefore, to obtain the limiting distribution of $N_e$, we just need to analyze $\tneb$, which is what we are going to do next.
\begin{figure}
\begin{centering}
\subfloat[\label{fig:LOO_quad}]{\begin{centering}
\includegraphics[scale=0.4]{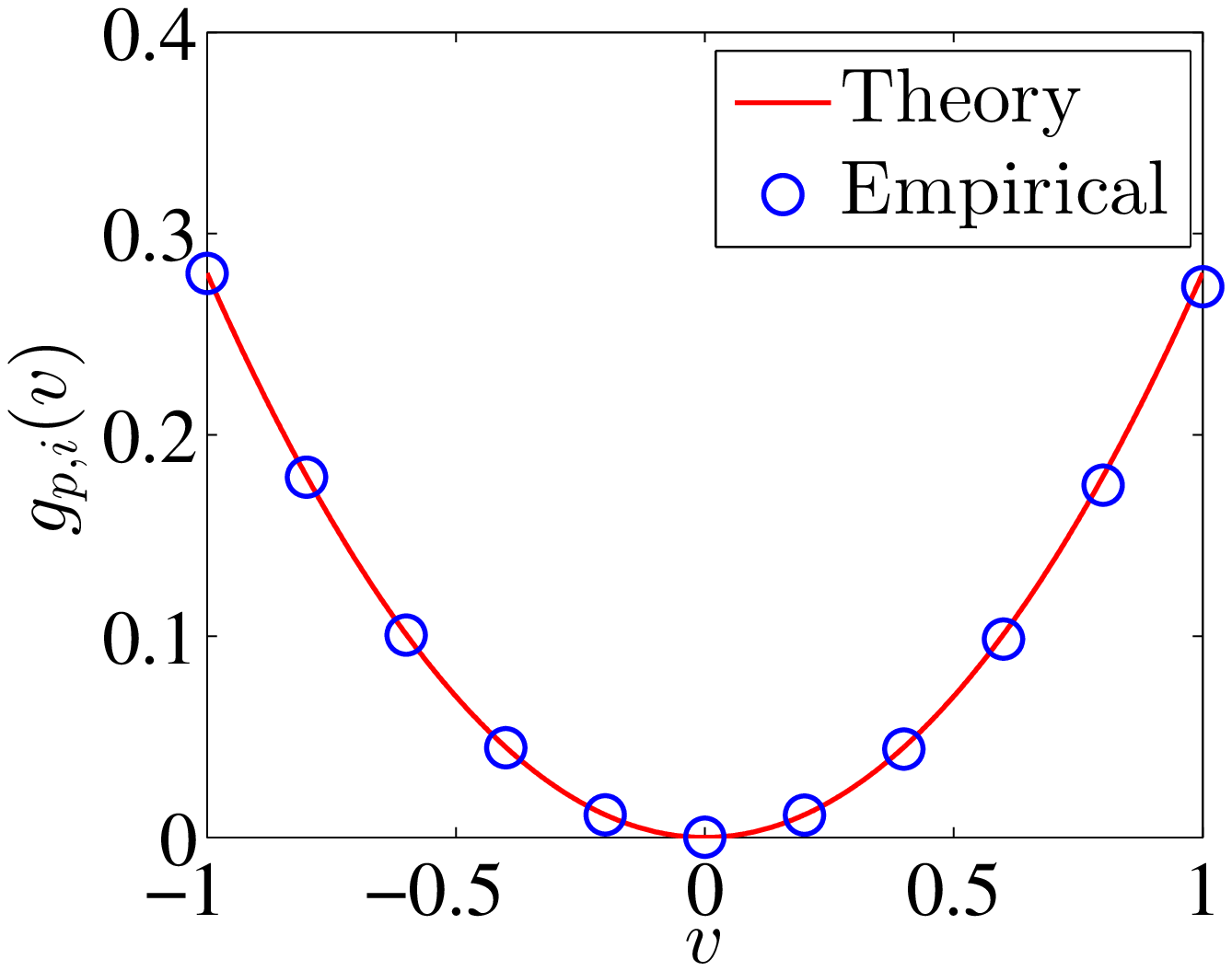}
\par\end{centering}
}\hphantom{}\subfloat[\label{fig:LOO_approx}]{\begin{centering}
\includegraphics[scale=0.4]{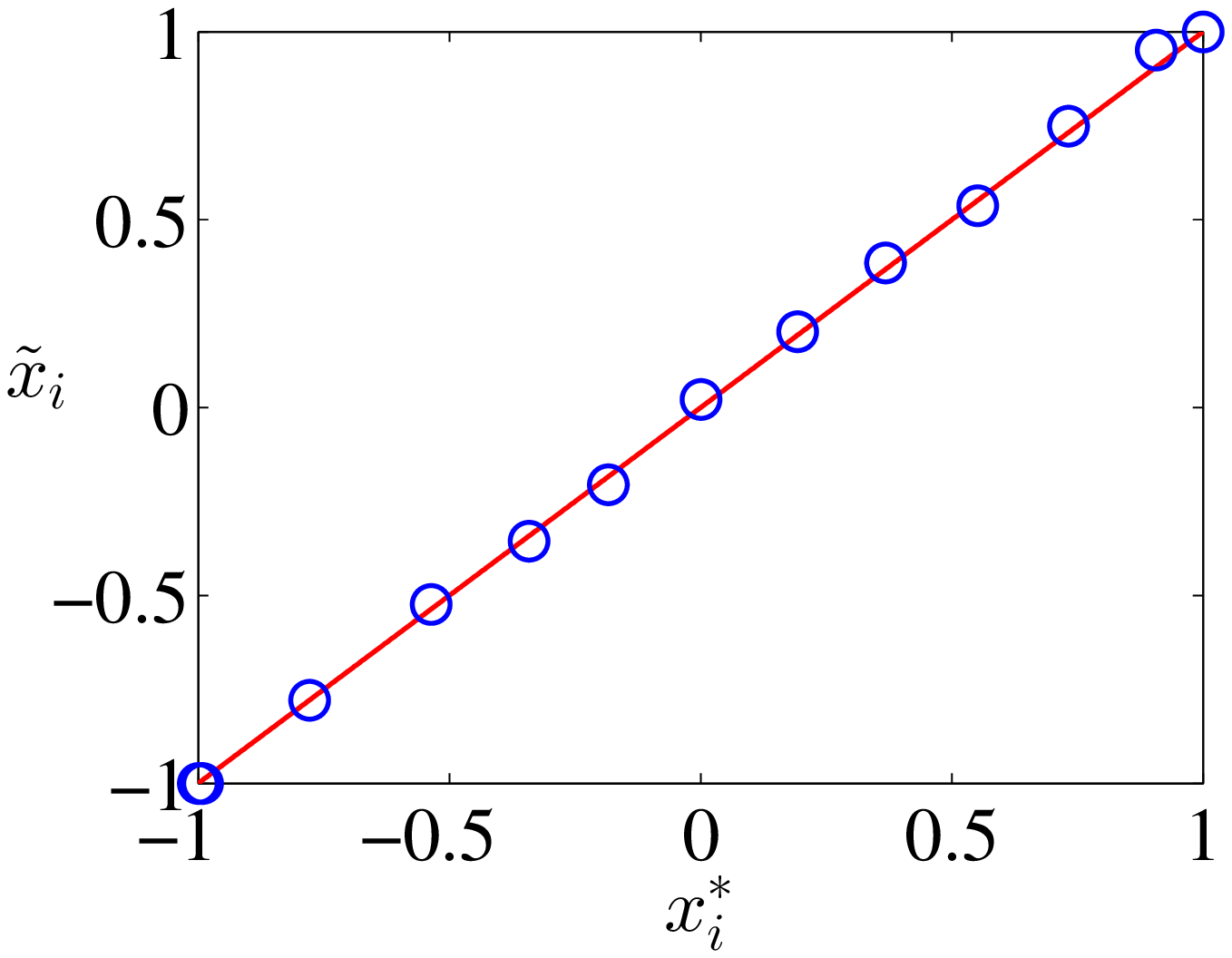}
\par\end{centering}
}
\par\end{centering}
\caption{\label{fig:LOO} Accuracy of the leave-one-out approximation. (a) Comparison of $g_{p,i}(v)$ with its limiting prediction $\frac{1}{2}A_p^{*}v^2$,
(b) Comparison between $\optxi{i}$ and its leave-one-out approximation $\xtdcav_i$.
In our experiments, $\nsdp^2 = 1$, $\sampratio = 1$ and $p = 1000$.}
\end{figure}
\subsection{Approximate independence of $\{\widetilde{\sgli}_i\}_{i\in[p]}$}
To derive the distribution of $\tneb$, we need to know the joint distribution of $\{\xtdcav_i\}_{i\in[p]}$. From  (\ref{eq:optxapprox2}), we know $\{\xtdcav_i\}_{i\in[p]}$ is determined by $\{\va_i^{\T}\vu_{\backslash i}^{*}\}_{i\in[p]}$.
Since for $i\neq j$, $\vu_{\backslash i}^{*} \approx \vu_{\backslash j}^{*}$, the set of variables $\{\xtdcav_i\}_{i\in[p]}$ are correlated, but the correlations are weak. In fact, we can prove something stronger. The following result, proved in \sref{tildex}, shows that any size-$k$ subset of $\{\va_i^{\T}\vu_{\backslash i}^{*}\}_{i\in[p]}$ are approximately independent, provided that $k$ is not too large.
\begin{prop}
\label{prop:approx_indep_Gaussian}
If $k\leq\sqrt{p},$ then $\text{there exists } c>0$ such that,
for any $b_{i}\in\R, i=1,2,\ldots,k$ and $\veps>0$,
\begin{equation}
\P\left(\bigcap_{i=1}^{k}\left\{ \va_{i}^{\T}\vu_{\backslash i}^{*}\leq b_{i}\right\} \right)\in\left[\prod_{i=1}^{k}\Phi\left(\frac{b_{i}-\sqrt{\sampratio}\veps}{\optS}\right)-{\Delta}_{p,k},\prod_{i=1}^{k}\Phi\left(\frac{b_{i}+\sqrt{\sampratio}\veps}{\optS}\right)+{\Delta}_{p,k}\right],\label{eq:joint_k_approx_indep_Gaussian}
\end{equation}
where $\Phi(\cdot)$ is the CDF of the standard Gaussian and ${\Delta}_{p,k}\bydef ckp^{\tfrac{1}{2}}e^{-c^{-1}p\min\left\{ \tfrac{\veps^{2}}{k^{2}},\tfrac{\veps}{\sqrt{p}}\right\} }$.
\end{prop}

It follows from (\ref{eq:optxapprox2}) and (\ref{eq:proxy_beta_hat}) that
$\big\{ \widetilde{\sgli}_{i}\neq\beta_{i}\big\} = \big\{ \va_{i}^{\T}\vu_{\backslash i}^{*}\leq-A_{p}^{*}\big\} $. (Recall that we have assumed that $\beta_i = -1$ for all $i$.)
By taking $b_{i}=-A_{p}^{*}$ in (\ref{eq:joint_k_approx_indep_Gaussian}),
we can conclude that the $k$ events $\{ \widetilde{\sgli}_{i}\neq\beta_{i}\} _{i\in[k]}$ (or equivalently $\{\I_{\widetilde{\sgli}_{i}\neq\beta_{i}}\}_{i\in[k]}$)
are also approximately independent. This is made precise by the following proposition, whose proof can be found in Appendix \ref{appen:ApproximateIndependence}.
\begin{prop}
\label{prop:joint_k_Bern_approx_indep_1}If
$k\leq p^{\tfrac{1}{8}}$, $\text{there exists } c>0,$ such that
\begin{equation}
\P\left(\bigcap_{i=1}^{k}\left\{ \widetilde{\sgli}_{i}\neq\beta_{i}\right\} \right)\in\left[\Phi^{k}\left(-\tfrac{1+cp^{-{1}/{4}}}{\opttau}\right)-ce^{-p^{1/4}/c},\Phi^{k}\left(-\tfrac{1-cp^{-{1}/{4}}}{\opttau}\right)+ce^{-p^{1/4}/c}\right].\label{eq:joint_k_Bern_approx_indep_2}
\end{equation}
Moreover, if $\nsdp^2\geq \frac{c'}{\log^2 p}$ for some $c' > 0$, then for all large enough $p$, 
\begin{equation}
\Big|\P\Big(\bigcap_{i=1}^{k}\left\{ \widetilde{\sgli}_{i}\neq\beta_{i}\right\} \Big)-\Phi^{k}\left(-\tfrac{1}{\opttau}\right)\Big|\leq\Phi^{k}\left(-\tfrac{1}{\opttau}\right)kp^{-{1}/{4}}\polylog p,\label{eq:joint_k_Bern_approx_indep_1}
\end{equation}
\end{prop}

\subsection{Proof of the Main Theorem}
\label{sec:mainthmproof}

We are now ready to prove Theorem~\ref{thm:PT_limitlaw_Ne} by showing that the limiting distribution of $\tneb$ converges to Poisson. Recall that $\tneb = \sum_{i=1}^{p} \I_{\widetilde{\sgli}_{i}\neq\beta_{i}}$. The approximate independence of  $\{\I_{\widetilde{\sgli}_{i}\neq\beta_{i}}\}$ makes the analysis tractable. Classical results on Poisson approximation of rare events deal with the sum of $p$ i.i.d. Bernoulli random variables with success probability $\lambda / p$. As $p \to \infty$, the sum converges in distribution to a Poisson random variable with rate $\lambda$. Things are slightly different in
our case, since $\tneb$ is a summation of $p$
weakly correlated Bernoulli random variables. The following proposition, proved in \sref{tneb_Poisson}, shows that the Poisson convergence still holds under the weaker condition of approximate
independence.
\begin{prop}
\label{prop:Poisson_convergence}
If $\limsup_{p\to\infty} \tfrac{\optlam}{\sqrt{\log p}} < \infty$, then
\begin{equation}
d_{\text{TV}}(\tneb,\mathscr{P}(\lambda_{p})) \le p^{-1/5}\polylog p,\label{eq:Poisson_convergence_TV}
\end{equation}
where $\mathscr{P}(\lambda)$
denotes a Poisson distribution with parameter $\lambda$.
\end{prop}

Finally, since the TV distance is a metric, the statement of Theorem~\ref{thm:PT_limitlaw_Ne} immediately follows from (\ref{eq:dTV_tNe_Ne_bd_1}), (\ref{eq:Poisson_convergence_TV}) and the triangle inequality.

\subsection{Proof of Proposition~\ref{prop:PT}}
\label{sec:PT}

Using the Gaussian tail bounds (\ref{eq:Phi_1_tau_upbd}) and (\ref{eq:Phi_1_tau_lowerbd}) given in Appendix \ref{appen:mills}, we can get
\begin{equation}
\label{eq:optlaminf}
\lim_{p\to\infty} \optlam =
\begin{cases}
0, &~\limptline > 1,\\
\infty, &~\limptline < 1.
\end{cases}
\end{equation}
Therefore, if $\limptline > 1$, it directly follows from Theorem \ref{thm:PT_limitlaw_Ne} that $\P(\neb = 0) = 1$.

The case that $\alpha^\ast < 1$ is more complicated. One can show that $\lambda_p\geq p^{c(\alpha^\ast)}$, where $c(\alpha^\ast)$ is some constant, so it is possible $\lim_{p\to\infty}d_{\text{TV}}(\tneb,\neb)\not\to0$.
Instead,
we can look at a subset $\mathcal{K}\subset[p]$. Define 
$N_{e,\mathcal{K}}$ as
the number of error bits in $\mathcal{K}$.
and $\lambda_{p,\mathcal{K}} \bydef {|\mathcal{K}|\Phi(-{\opttau^{-1}})}$.
We can find
$\mathcal{K}$ satisfying $ \lambda_{p,\mathcal{K}} \asymp \sqrt{\log p}$.
Then following same steps of proving Proposition \ref{prop:TVconvergence} and Proposition~\ref{prop:pt_tildeNEB} in Appendix \ref{subsec:pt_tildeNEB}, we can show 
$\lim_{p\to\infty}\P(N_{e,\mathcal{K}}=0)=0$, which indicates that
$\lim_{p\to\infty}\P(\neb=0)=0$, since $N_{e,\mathcal{K}} \leq \neb$.

Finally, we prove (\ref{eq:Ne_2ndOrder_transition}). If $\ptline$ satisfies (\ref{eq:alphap_nearPT}), then for large $p$, $\nsdp^2 \asymp (\log p)^{-1}$. Letting $t=\nsdp^2$ in \eqref{eq:tau_t_equation}, it follows that if $\ptline\to \alpha^{*}$, then $2\ptline\opttau^2\log p \to 1$. On the other hand, from the auxiliary bounds (\ref{eq:mills_ratio}) given in Appendix~\ref{appen:mills}, we can get $ \frac{m(-\opttau^{-1})}{\opttau}\to 1$. Applying (\ref{eq:lambda}) and (\ref{eq:TV_inf}) gives us
\begin{align}
\lim_{p\to\infty}\P(\neb=0) &= \lim_{p\to\infty} \exp\left\{ -p\Phi(-{1}/{\opttau}) \right\} \nonumber\\
&\teq{\text{(a)}} \lim_{p\to\infty} \exp\left\{ -p \cdot \opttau \varphi(-{1}/{\opttau}) \right\} \nonumber\\
&\teq{\text{(b)}} \lim_{p\to\infty} \exp\left\{ -p (2\ptline\log p)^{-1/2} \frac{e^{-\ptline \log p}}{\sqrt{2\pi}} \right\} \nonumber\\
&= \lim_{p\to\infty} \exp\left\{ -\exp\left\{ -\log p \left( \ptline - 1 + \frac{\log(\ptline)}{2\log p} + \frac{\log (4\pi)+\log\log p}{2\log p} \right) \right\}  \right\}\nonumber\\
&\teq{\text{(c)}} e^{-e^{-x}},\nonumber
\end{align}
where step (a) follows from $ \frac{m(-\opttau^{-1})}{\opttau}\to 1$, step (b) follows from $2\ptline\opttau^2\log p\to 1$ and we use (\ref{eq:alphap_nearPT}) in step (c).

\section{The Limiting Quadratic Function}\label{sec:completeproof}

The goal of this technical section is to make the approximations shown in \fref{LOO} rigorous.

\subsection{Proof of Proposition \ref{prop:gp_conv_2}}
\label{sec:scalarCP}
To lighten notation, we will sometimes omit the leave-one-out subscript as used in Sec. \ref{sec:ScalarProblemOV}. For example, $\dmtx_{\backslash i}$ will be replaced by $\dmtx$, and $\va_i$ by $\vacav$, as long as doing so causes no confusion.

Let us first introduce the following function:
\begin{align}
\mathcal{G}_{p}(\vs) & \bydef\max_{\vu}[\vs^{\T}\vu-\Lcav(\vu)]-[\vs^{\T}\vu^{*}-\Lcav(\vu^{*})],\label{eq:Gpx},
\end{align}
where $\Lcav(\vu) = \|\dmtxcav^{\T}\vu\|_{1}+\vu^{\T}\vycav+\frac{1}{2}\|\vu\|^{2}$ and $\vucav^{*}=\argmin{\vu}\Lcav(\vu)$.
Using $\mathcal{G}_{p}(\vs)$ and omitting subscript $i$, scalar function $g_{p,i}(v)$ defined in \eqref{eq:L_tilde} can be also expressed as:
\begin{equation*}
g_{p}(v)  = \mathcal{G}_{p}(\vacav v)   
\end{equation*}
and correspondingly, we re-write (\ref{eq:scalar_problem_1a}) as:
\begin{equation}
\min_{-1\leq \xcav \leq1}g_{p}(\xcav-\sglicav)+\vacav^{\T}\vu^{*}(\xcav-\sglicav).\label{eq:scalar_problem_1}
\end{equation}
%
It can be seen that $\mathcal{G}_{p}(\vs)$ is related with the conjugate function of $\Lcav(\vu)$, which is a strongly convex function. Therefore, $\mathcal{G}_{p}(\vs)$ and $g_{p}(v)$ possess some nice properties that will be useful in our proof. We gather them together in Appendix \ref{appendix:Gpsgpv}.

We first show that $g_{p}(v)$ concentrates around its expectation, which is the following proposition. Its proof will be given in Appendix \ref{appen:uniformconv}.
\begin{prop}
\label{prop:uniform_convergence} $\text{ There exists } c>0$, s.t. for any $\varepsilon>0$,
\begin{equation}
\P\left(\sup_{v\in[-2,2]}\left|g_{p}(v)-\E g_{p}(v)\right|>\varepsilon\right)\leq \frac{c\sampratio}{\veps}e^{-c^{-1}p\min\left\{ \tfrac{\veps^{2}}{\sampratio},\veps\right\} }.\label{eq:concentration_gp_uniform}
\end{equation}
\end{prop}

The next result shows that $\E g_{p}(v)$
is essentially a quadratic function in the large $p$ limit.
\begin{prop}
\label{prop:quad1} For any $v\in[-2,2]$,
\begin{equation}
\left|\E g_{p}(v)-\frac{1}{2}A_{p}v^{2}\right|\leq\frac{16\sampratio}{\nsdp^{2}p},\label{eq:quad1}
\end{equation}
where
$A_p$ is defined in (\ref{eq:Ap}).
\end{prop}
\begin{IEEEproof}
First we introduce the following auxiliary functions:
\begin{align}
\losspo p(\field)&\bydef\min_{\vx\in[-1,1]^{p}}\frac{\|\mA\vx-\vy+\sqrt{\field}\tvacav\|^{2}}{2p},~\field \geq 0,\label{eq:Qp_theta_def}
\end{align}
where $\tvacav \sim \mathcal{N}(0,\mI_n)$, independent of $\dmtx, \noise$. Clearly, the original problem \eqref{eq:box_LASSO} is the special case when $\field=0$. For notational convenience, we also define the expectation of $\losspo p(\field)$ as:
\begin{align}
\averlosspo p(\field) &\bydef \E {\losspo p(\field)}\nonumber \\
&= \frac{1}{p}\E\max_{\vu}\vu^{\T}\sqrt{\field}\tvacav -L(\vu), \label{eq:averQp_def}
\end{align}
where $L(\vu)$ is given in \eqref{eq:Gpx}. Note that the connection between $\averlosspo p(\field)$ and $\E g_{p}(v)$ is:
\begin{equation}
\E g_{p}(v)=\frac{\averlosspo {p}({v^2}/{p})-\averlosspo {p}(0)}{{v^2}/{p}}v^2,\label{eq:avergp_averGp_relation}
\end{equation}
i.e., $\E g_{p}(v)$ can be approximated by the derivative
of $\averlosspo p(\field)$ at $\field=0$. To make this intuition rigorous,
we need to study the analytical properties of $\averlosspo p(\field)$.

First, we show that $\averlosspo p(\field)$ is
differentiable on $[0,\infty)$ and $\averlosspo p'(\field)$
is Lipschitz continuous. Indeed, from (\ref{eq:Gpx}) and (\ref{eq:averQp_def}),
\begin{align}
\averlosspo p'(\field)
 & \teq{}\frac{1}{p}\frac{\partial}{\partial \field}\left[\E\max_{\vu}\vu^{\T}\sqrt{\field}\tvacav -L(\vu)\right]\nonumber \\
 & =\frac{1}{p}\frac{\partial}{\partial \field}\E\max_{\vu}-\left(\|\dmtxcav^{\T}\vu\|_{1}+\sglcav^{\T}\dmtxcav^{\T}\vu+\frac{1}{2}\|\vu\|^{2}+\sqrt{\field+\nsdp^{2}}\vu^{\T}\widetilde{\noise}\right)\label{eq:averGp_opt2}\\
 & \teq{\text{(a)}}-\frac{\E\widetilde{\noise}^{\T}\hat{\vu}_{\field}}{2p\sqrt{\field+\nsdp^{2}}},\label{eq:averGp_grad2}
\end{align}
where $\widetilde{\noise}\sim\mathcal{N}(\boldsymbol{0},\mI_{n})$
and $\hat{\vu}_{\field}$ corresponds to the optimal solution of (\ref{eq:averGp_opt2}).
In step (a), we use dominated convergence theorem (DCT) to interchange derivative and expectation.
By the same argument of (\ref{eq:ux_Lipschitz}) in Appendix \ref{appendix:Gpsgpv}, we have for any $b,c \geq 0$,
\begin{equation}
\|\hat{\vu}_{b}-\hat{\vu}_{c}\|\leq\left|\sqrt{b+\nsdp^{2}}-\sqrt{c+\nsdp^{2}}\right|\|\widetilde{\noise}\|.\label{eq:Lipschitz_uh}
\end{equation}
On the other hand, for any $\field \geq 0$,
\begin{equation}
\label{eq:bdnorm_uh}
\|\hat{\vu}_{\field}\| = \min_{\vx\in[-1,1]^{p}}\|\dmtxcav \vxcav - (\dmtxcav \sglcav+\sqrt{\field+\nsdp^{2}}\widetilde{\noise})\| \leq \sqrt{\field+\nsdp^{2}}\|\widetilde{\noise}\|.
\end{equation}
Combining \eqref{eq:averGp_grad2}, \eqref{eq:Lipschitz_uh} and \eqref{eq:bdnorm_uh}, for any
$b>c\geq0$, we can get
\begin{align}
\left|\averlosspo p'(b)-\averlosspo p'(c)\right|
\leq \frac{\sampratio|b-c|}{\nsdp^{2}}.\label{eq:Lipschitz_averG_deri}
\end{align}
Therefore, $\averlosspo p'(h)$ is $\frac{\sampratio}{\nsdp^{2}}$-Lipschitz.

Now we are ready to analyze $\E g_{p}(v)$. By the mean value theorem,
 we get from (\ref{eq:avergp_averGp_relation}) that
\begin{align}
\E g_{p}(v) & =\averlosspo p'\left(\tfrac{\kappa_{p}v^2}{p}\right)v^2,\label{eq:avergp_2}
\end{align}
where $\kappa_{p}\in[0,1]$. From (\ref{eq:Lipschitz_averG_deri}) and
(\ref{eq:avergp_2}), we deduce that
\begin{equation}
\left|\E g_{p}(v)-\averlosspo p'(0)v^2\right|\leq
\frac{v^4\sampratio}{\nsdp^{2}p}\leq\frac{16\sampratio}{\nsdp^{2}p}.\label{eq:Lipschitz_averG_deri_2}
\end{equation}
On the other hand, from (\ref{eq:averGp_grad2}),
\begin{equation}
\averlosspo p'(0)=-\frac{\E\widetilde{\noise}^{\T}\hat{\vu}_{0}}{2\nsdp^{2}p}=-\frac{\E\noise^{\T}\optu }{2\nsdp^{2}p},\label{eq:Lipschitz_averG_deri0}
\end{equation}
It can be checked from (\ref{eq:optu_dual}) that $\vucav^{*} = \dmtxcav\optx - \vycav$. Combining (\ref{eq:Lipschitz_averG_deri_2}) and (\ref{eq:Lipschitz_averG_deri0}),
we get (\ref{eq:quad1}).
\end{IEEEproof}
\begin{rem}
It will be shown later [c.f. (\ref{eq:Apstar_bd})] that $A_p\geq C\sampratio$, for some constant $C>0$. Therefore, we know from (\ref{eq:quad1}) that the quadratic approximation of $\E g_{p}(v)$
is accurate for large $p$, if $\nsdp\gg p^{-1/2}$. We will prove that, when $\nsdp<\frac{c}{\sqrt{\log p}}$
for some constant $c$, perfect recovery is achieved with high probability.
This means that $\nsdp\gg p^{-1/2}$ already covers the regime where we are most interested in. In the following, we will take $\nsdp\geq \frac{1}{\log p}$.
\end{rem}

Proposition \ref{prop:uniform_convergence} and \ref{prop:quad1} immediately implies the first part of Proposition \ref{prop:gp_conv_2}, i.e., (\ref{eq:gp_conv_2}). Next we show $A_{p}$ converges to $A_{p}^{*}$ in the high-dimensional limit.
From \eqref{eq:Lipschitz_averG_deri0},
\begin{equation}
A_{p}={2\averlosspo p'(0)}.\label{eq:Ap_3}
\end{equation}
Hence, it boils down to analyzing $\averlosspo p'(\field)$ and its limit, which can be done as follows.

\subsubsection{Convergence of $\losspo p(\field)$}
The CGMT framework in \cite{thrampoulidis2018symbol,miolane2018distribution} can be readily applied to computing the limit
of $\losspo p(\field)$ in high dimensions.
\begin{lem}
\label{lem:Qp_theta_concentration}
$\text{ There exists } c>0$, s.t., for any $\varepsilon>0$ and $\field\in[0,1]$,
\begin{equation}
\P\left(|\losspo p(\field)-\optQ(\field)|>\veps\right)\leq
\tfrac{ce^{-p\min\big\{\tfrac{\veps^2}{\sampratio }, \veps\big\}/c}}{\min\{\tfrac{\veps}{\sampratio},\sqrt{\tfrac{\veps}{\sampratio}}\}},\label{eq:Qp_theta_concentration}
\end{equation}
where
\begin{equation}
\optQ(\field)=\frac{1}{2}\left[\min_{\tau>0} F_{p}\big(\tau;\field+\nsdp^2,\sampratio\big)   \right]^{2},\label{eq:Qpstar_theta}
\end{equation}
with $F_{p}$ defined in (\ref{eq:F}). Also for any $\gamma>2$, $\text{ there exists } c>0$ such that
\begin{equation}
\sup_{\field\in[0,1]}|\averlosspo p(\field)-\optQ(\field)|<cp^{-1/\expnt}.\label{eq:EQp_theta_concentration}
\end{equation}
\end{lem}
\begin{rem}
The proof of Lemma \ref{lem:Qp_theta_concentration} will be given in Appendix \ref{appendix:CGMTproof}. We can find $\optQ(0) = \frac{\optS^2}{2}$, where $\optS$ is defined in (\ref{eq:tau}). This can be understood from (\ref{eq:Qp_theta_def}) and (\ref{eq:Qp_theta_concentration}), since $\optQ(\field)$ is the limiting value of the squared fitting error when the noise variance is $\field + \nsdp^2$.
\end{rem}
\subsubsection{Smoothness of $\optQ(\field)$}
\begin{lem}
\label{lem:Qconvexity}
$\optQ(\field)$
is twice differentiable over $\field\geq 0$, with
\begin{equation}\label{eq:Qpstar_deri}
\optQ\;'(0) = \frac{\optS}{2\opttau}
\end{equation}
and $\optQ\,''(\field)\leq C$,
for all $\field\geq 0$, where $C$ is some constant.
\end{lem}
\begin{IEEEproof}
Note that $\optQ(\field)$ is a composition of $R_p(t)$ and $t(\field) = \field + \nsdp^2$, where $R_p(t)$ is defined in Appendix \ref{appen:1Dopt}. By chain rule, $\optQ(\field)$ is twice differentiable, with $\optQ\;'(0) = \tfrac{\optS}{2\opttau}$ and
\begin{align}
\label{eq:DDQp_theta_ubd}
Q_{p}^{*}\,''(\field) & =R_{p}''(t)t'(\field)+R_{p}'(t)t''(\field)\nonumber \\
&= {R_{p}''(\field+\nsdp^2)}.
\end{align}
Then together with bound \eqref{eq:Rp_2ndDeri_bd} shown in Appendix \ref{appen:1Dopt}, we know there exists $C>0$, s.t., $\optQ\,''(\field)\leq C$,
for all $\field\geq 0$.
\end{IEEEproof}
\subsubsection{Convergence of $A_{p}$ to $A_{p}^{*}$}
Now we can show the convergence of the curvature $A_{p}$, which also implies the simple limiting form of $g_{p}(v)$.
\begin{lem}
There exists $c>0$ such that
\begin{equation}
|A_{p}-A_{p}^{*}|<cp^{-1/(2\expnt)}.\label{eq:A_p_conv}
\end{equation}
\end{lem}
\begin{IEEEproof}
For $\expnt>2$, there exists $C>0$, s.t. for $\field\in(0,1]$,
\begin{align}
|A_{p} - A_{p}^{*}| &\tleq{(\text{a})} 2\left| \averlosspo {p}'(0) - \tfrac{\averlosspo {p}(\field)-\averlosspo {p}(0)}{\field} \right|
+ 2\left| \tfrac{\averlosspo {p}(\field)-\averlosspo {p}(0)}{\field} - \tfrac{\optQ(\field)-\optQ(0)}{\field}\right|\nonumber\\
&~~ + 2\left| \tfrac{\optQ(\field)-\optQ(0)}{\field} - \optQ\;'(0) \right|\nonumber\\
&\tleq{(\text{b})} C\left( \tfrac{\field\sampratio}{\nsdp^2} + \tfrac{\sqrt{\sampratio}p^{-\tfrac{1}{\expnt}}}{\field} + \field \right),
\end{align}
where in step (a), we use \eqref{eq:Apstar}, \eqref{eq:Ap_3} and \eqref{eq:Qpstar_deri} and in step (b), we use \eqref{eq:Lipschitz_averG_deri}, \eqref{eq:EQp_theta_concentration} and Lemma \ref{lem:Qconvexity}. Therefore, taking $\field = p^{-\tfrac{1}{2\expnt}}$ and using Assumptions \ref{asmp:A3} and \ref{asmp:A4}, we can get \eqref{eq:A_p_conv}.
\end{IEEEproof}
%
\subsection{Proof of Proposition \ref{prop:approximation_xi_xtildei}}
\label{sec:convoptxCP}


Proposition \ref{prop:gp_conv_2} indicates that the original scalar problem
(\ref{eq:scalar_problem_1}) can be well approximated by
\begin{align}
 & \min_{\xcav\in[-1,1]}\frac{1}{2}A_{p}(\xcav-\sglicav)^{2}+\vacav^{\T}\vucav^{*}(\xcav-\sglicav),\label{eq:scalar_problem_2}
\end{align}
which has an {explicit} optimal solution:
\begin{equation}\label{eq:optxapprox1}
\xhatcav=\tprox_{[-1,1]}\left(\sglicav-\frac{\vacav^{\T}\vucav^{*}}{A_{p}}\right).
\end{equation}
Note that the difference between $\xhatcav$ and $\xtdcav$ should be small, as implied by \eqref{eq:optxapprox2}, \eqref{eq:optxapprox1} and \eqref{eq:A_p_conv}. In fact, we can directly prove $\xtdcav \to \optxcav$ without considering $\xhatcav$. The reason for us to introduce this intermediate variable is to achieve a better convergence rate in our proof.

The first lemma below shows that the objective function of (\ref{eq:scalar_problem_2}), i.e.,
\begin{equation}
\widehat{\ell}_{p} (\xcav) = \frac{1}{2}A_{p}(\xcav-\sglicav)^{2}+\vacav^{\T}\vucav^{*}(\xcav-\sglicav)
\end{equation}
is strongly convex.
\begin{lem}
$\text{There exists } K>0$, s.t., $A_{p}\geq K\sampratio$ for all $p$ large enough. Therefore, $\widehat{\ell}_p (\xcav)$ is $K\sampratio$-strongly convex.
\end{lem}
\begin{IEEEproof}
By (\ref{eq:tau}) and the definition of $A_{p}^{*}$, we have
\begin{align}
\label{eq:Apstar_bd}
A_{p}^{*} & =\frac{1}{2}\left(\sampratio-\frac{1}{2}\right)+\frac{\nsdp^{2}}{2\opttau^{2}}+\frac{1}{2}\int_{\frac{2}{\opttau}}^{\infty}\left(x-\frac{2}{\opttau}\right)^{2}\Phi(dx) \geq\frac{1}{2}\left(\sampratio-\frac{1}{2}\right).
\end{align}
Then from assumption \ref{asmp:A4} and (\ref{eq:A_p_conv}), we
know there exists $K>0$ s.t. $A_{p}\geq K\sampratio>0$ and $\widehat{\ell}_{p}(x)$ is $K\sampratio$-strongly convex.
\end{IEEEproof}
Then together with uniform convergence proved in Proposition \ref{prop:gp_conv_2}, we can show $\optxcav\to\xhatcav$.
\begin{lem}
\label{lem:conv_xistar2xitilde}$\text{There exists } c>0$
s.t., for $\varepsilon\in(0,1)$,
\begin{equation}
\P\left(\left|\optxcav -\xhatcav\right|>\veps\right)<\frac{c}{\veps^{2}}e^{-p\veps^{4}/c}.\label{eq:conv_xistar2xitilde}
\end{equation}
\end{lem}
\begin{IEEEproof}
Since $\widehat{\ell}_{p}(x)$ is $K\sampratio$-strongly convex,
\begin{equation}
\widehat{\ell}_{p}(\optxcav)-\widehat{\ell}_{p}(\xhatcav)\geq\frac{1}{2}K\sampratio(\optxcav -\xhatcav)^{2}.\label{eq:strong_convexity_2}
\end{equation}
Let $\ell_p(x)$ be the objective function in \eqref{eq:scalar_problem_1a}. From (\ref{eq:gp_conv_2}) we know $\text{there exists } c>0$, s.t., for $\veps\in(0,1)$, $|\widehat{\ell}_{p}(\optxcav)-\ell_{p}(\optxcav)|\leq{\sampratio}\veps$
and $|\widehat{\ell}_{p}(\xhatcav)-\ell_{p}(\xhatcav)|\leq{\sampratio}\veps$
with probability greater than $1-\frac{c}{\veps}e^{-p\veps^{2}/c}$.
This indicates
\begin{equation}
\widehat{\ell}_{p}(\optxcav)-\hat{\ell}_{p}(\xhatcav)\leq[\ell_{p}(\optxcav)+\sqrt{\sampratio}\veps]-[\ell_{p}(\xhatcav)-\sqrt{\sampratio}\veps]\leq 2{\sampratio}\veps.\label{eq:minimum_condition_2}
\end{equation}
From (\ref{eq:strong_convexity_2}) and (\ref{eq:minimum_condition_2}),
we can get $\text{there exists } c>0$ s.t. for all $\veps\in(0,1)$, $\P\left(\left|\optxcav-\xhatcav\right|>\sqrt{\veps}\right)<\frac{c}{\veps}e^{-p\veps^{2}/c}$.
Then changing $\sqrt{\veps}$ to $\veps$ in the above, we get (\ref{eq:conv_xistar2xitilde}).
\end{IEEEproof}
Furthermore, using (\ref{eq:A_p_conv}) we can also show $\xhatcav\to\xtdcav$.
\begin{lem}
\label{lem:conv_xitilde2xitilde_star}For $\expnt>2$, $\text{there exists } c>0$,
s.t., for $\varepsilon\in(0,1)$,
\[
\P\left(\left|\xhatcav-\xtdcav \right|>\veps\right)<\frac{c}{\veps}e^{-p^{\tfrac{1}{\expnt}}\veps^{2}/c}.
\]
\end{lem}
\begin{IEEEproof}
By the non-expansiveness of proximal operator $\tprox_{[-1,1]}(\cdot)$, from \eqref{eq:optxapprox2} and  \eqref{eq:optxapprox1} we know there exists $C>0$, s.t.,
\begin{equation}\label{eq:xhat_xtd_close}
\left|\xhatcav-\xtdcav\right|\leq\left|\frac{1}{A_{p}}-\frac{1}{A_{p}^{*}}\right||\vacav^{\T}\vucav^{*}|\leq \frac{C}{\sampratio^2}|\vacav^{\T}\vucav^{*}|p^{-\tfrac{1}{2\expnt}},
\end{equation}
where we have used (\ref{eq:A_p_conv}) and (\ref{eq:Apstar_bd}).
Recall that $\vucav^{*} = \dmtxcav\optx - \vycav$, so similar to (\ref{eq:sqrt_Qpstar_convergence}) and (\ref{eq:sqrt_Qpstar_convergence2}), we obtain that $\text{there exists } c>0$,
s.t., for all $\veps>0$, $\P\left(\left|\tfrac{\|\vucav^{*}\|}{\sqrt{p}}-\optS\right|>\veps\right)\leq \tfrac{c\sqrt{\sampratio}}{\veps}e^{-p\veps^{2}/c}.$
Since $\vacav$ and $\vucav^{*}$ are independent, then from (\ref{eq:xhat_xtd_close}) it is not hard to show $\text{there exists } c>0$, s.t., for all $\varepsilon\in(0,1)$,
$\P\left(\left|\xhatcav-\xtdcav\right|>\veps\right)\leq\tfrac{c}{\veps}e^{-p^{1/\expnt}\veps^{2}/c}$.
\end{IEEEproof}
Lemma \ref{lem:conv_xistar2xitilde} and \ref{lem:conv_xitilde2xitilde_star}
imply Proposition \ref{prop:approximation_xi_xtildei}, based on which we can now prove Proposition \ref{prop:TVconvergence}.
\subsection{Proof of Proposition \ref{prop:TVconvergence}}
\label{sec:convoptxCP1}
Our strategy is to show that $\P(\widetilde{\sgl}\neq\hat{\sgl})$
is small, which implies $\P(\tneb\neq\neb)$ is small and so is $d_{\text{TV}}(\tneb,\neb)$. Recall that $\tneb$ and $\neb$ are in the same probability space, and we have assumed $\sgli_{i}=-1$, for any $i\in[p]$.
Then the following simple relation holds:
\begin{equation}
\big\{ \widetilde{\sgli}_{i}\neq\hat{\sgli}_{i}\big\} \subset\big\{ |\tilde{x}_{i}-\optxi i|>p^{-\tfrac{1}{5}}\big\} \bigcup\big\{ \tilde{x}_{i}\in[-p^{-\tfrac{1}{5}},p^{-\tfrac{1}{5}}]\big\} .\label{eq:betaneq_inclusion_1}
\end{equation}
Since $\tilde{x}_{i}=\tprox_{[-1,1]}\left(\sgli_{i}-\tfrac{\va_{i}^{\T}\vu_{\backslash i}^{*}}{A_{p}^{*}}\right)$,
for $U\in(-1,1)$, $|\tilde{x}_{i}|\leq U\Leftrightarrow\Big|\tfrac{\va_{i}^{\T}\vu_{\backslash i}^{*}}{A_{p}^{*}}+1\Big|\leq U$.
Then letting $b_{i}=A_{p}^{*}(-1+p^{-\tfrac{1}{5}})\text{ and }A_{p}^{*}(-1-p^{-\tfrac{1}{5}})$,
$k=1$ and $\veps=p^{-\tfrac{1}{5}}$ in (\ref{eq:joint_k_approx_indep_Gaussian}),
we can show $\P(|\tilde{x}_{i}|\leq p^{-\tfrac{1}{5}})\leq\Phi(-1/\opttau)p^{-\tfrac{1}{5}}\polylog p$,
similar to (\ref{eq:joint_k_approx_indep_Gaussian_4_1}) shown in Appendix \ref{appen:ApproximateIndependence}.
On the other hand, letting $\veps=p^{-\tfrac{1}{5}}$ in (\ref{eq:conv_xistar}),
$\P(\left|\optxi i-\tilde{x}_{i}\right|>p^{-\tfrac{1}{5}})<p^{\tfrac{2}{5}}e^{-p^{1/12}/c}$.
These together with (\ref{eq:betaneq_inclusion_1}) indicate
\begin{equation}
\P(\widetilde{\sgli}_{i}\neq\hat{\sgli}_{i})\leq\Phi(-1/\opttau)p^{-\tfrac{1}{5}}\polylog p.\label{eq:single_coordinate_notmatch}
\end{equation}
By union bound,
\[
\P(\widetilde{\sgl}\neq\hat{\sgl}) \leq \sum_{i=1}^{p}\P(\widetilde{\sgli}_{i}\neq\hat{\sgli}_{i})\leq\lambda_{p}p^{-\tfrac{1}{5}}\polylog p.
\]
Since $d_{\text{TV}}(\tneb,\neb)\leq\P(\tneb\neq\neb)\leq\P(\widetilde{\sgl}\neq\hat{\sgl})$,
we obtain (\ref{eq:dTV_tNe_Ne_bd_1}).


\section{Asymptotic Distributions}
\label{sec:surrogate}

This is another technical section. Our main goal here is to derive the asymptotic distribution of  $\set{\widetilde{x}_i}$ and that of $\tneb$.

\subsection{Proof of Proposition \ref{prop:approx_indep_Gaussian}}
\label{sec:tildex}

By the exchangeability of $\big\{ \va_{i}^{\T}\vu_{\backslash i}^{*}\big\} _{i\in[p]}$, we just need to consider the joint distribution of $\big\{ \va_{i}^{\T}\vu_{\backslash i}^{*}\big\} _{i\in[k]}$,
i.e., the first $k$ coordinates. A key result we are going
to establish is that $\big\{ \va_{i}^{\T}\vu_{\backslash i}^{*}\big\} _{i\in[k]}$
are approximately independent,
provided that $k$ is not too large.

Let $\vu_{\backslash[k]}^{*}$ be the optimal solution of
\begin{equation}
\label{eq:leave_k_out_Lu}
\min_{\vu}\|\dmtx_{\backslash[k]}^{\T}\vu\|_{1}+\vu^{\T}\dmtx_{\backslash[k]}\vbeta_{\backslash[k]}+\frac{1}{2}\|\vu\|^{2}+\vu^{\T}\vw,
\end{equation}
where $\dmtx_{\backslash[k]}$ is the matrix formed by removing the
first $k$ columns of $\dmtx$ and $\vbeta_{\backslash[k]}$ is defined
in the same way. In other words, $\vu_{\backslash[k]}^{*}$ is the
leave-$k$-out solution of $\min_{\vu} L(\vu)$. Also define
\begin{equation}
\widetilde{\vu}_{\backslash[k]}\bydef\frac{\sqrt{p}\optS\vu_{\backslash[k]}^{*}}{\|\vu_{\backslash[k]}^{*}\|}.\label{eq:utilde_star_notk}
\end{equation}
Since $\va_{i}\iid\mathcal{N}(\boldsymbol{0},\mI_{p}/p)$, $i=1,2,\ldots,k$ and $\widetilde{\vu}_{\backslash[k]}$
is independent of $\left\{ \va_{i}\right\} _{i\in[k]}$, with fixed
norm $\sqrt{p}\optS$, the joint distribution of $\left\{ \va_{i}^{\T}\widetilde{\vu}_{\backslash[k]}\right\} _{i\in[k]}$
is:
\begin{equation}
\begin{pmatrix}\va_{1}^{\T}\widetilde{\vu}_{\backslash[k]} & \va_{2}^{\T}\widetilde{\vu}_{\backslash[k]} & \ldots & \va_{k}^{\T}\widetilde{\vu}_{\backslash[k]}\end{pmatrix}^{\T}\sim\mathcal{N}(\boldsymbol{0},\optS^2\mI_{p}).\label{eq:joint_ai_utilde_exclude_k}
\end{equation}

Our proof of approximate independence of $\big\{ \va_{i}^{\T}\vu_{\backslash i}^{*}\big\} _{i\in[k]}$
consists of two steps:
\begin{enumerate}
\item Show the joint distribution of $\big\{ \va_{i}^{\T}\vu_{\backslash i}^{*}\big\} _{i\in[k]}$
is closed to that of $\big\{ \va_{i}^{\T}\vu_{\backslash[k]}^{*}\big\} _{i\in[k]}$.
This is proved in Lemma \ref{lem:leave1_leavek_compare}.
\item Show the joint distribution of $\big\{ \va_{i}^{\T}\vu_{\backslash[k]}^{*}\big\} _{i\in[k]}$
is closed to that of $\big\{ \va_{i}^{\T}\widetilde{\vu}_{\backslash[k]}\big\} _{i\in[k]}$,
which are mutually independent. This is proved in Lemma
\ref{lem:leavek_leaveknormalized_compare}.
\end{enumerate}
Details of the proof can be found in Appendix \ref{appen:ApproximateIndependence}.

\subsection{The Limiting Poisson Law of $\tneb$}
\label{sec:tneb_Poisson}


Before presenting the actual proof, it would help to first show some heuristic derivations.
We employ the following general inclusion-exclusion
principle \cite[p.106]{feller1968introduction}: for any $k\in[p]$, the
probability $P_{k}$ that exactly $k$ among $p$ events $A_{1},\ldots,A_{p}$
occur is
\begin{equation}
P_{k}=\sum_{m=k}^{p}\binom{m}{k}(-1)^{m-k}S_{m},\label{eq:Pk_def}
\end{equation}
where
\begin{equation}
S_{m}=\begin{cases}
1 & m=0,\\
\sum_{1\leq i_{1}<\cdots<i_{m}\leq p}\P\left(\bigcap_{j=1}^{m}A_{i_{j}}\right) & 1\le m\leq p.
\end{cases}\label{eq:Sk_def}
\end{equation}
In our setting, $A_{i}=\{ \tilde{\sgli}_{i}\neq\sgli_{i}\} ,i=1,2,\ldots,p$ and
$P_{k}=\P(\tneb=k)$.

By the exchangeability of $\{ \tilde{\sgli}_{i}\} _{i\in[p]}$, we have
$S_{m}=\binom{p}{m}S_{[m]},\label{eq:Sm_def_1},$
with $S_{[m]}=\P\left(\tilde{\sgli}_{i}\neq\sgli_{i},i\in[m]\right).$
From Proposition \ref{prop:joint_k_Bern_approx_indep_1},
for large enough $p$ and ``reasonably large'' $m$, $S_{[m]}\approx \Phi^{m}\left(-\tfrac{1}{\opttau}\right)$, so
\begin{align}
S_{m} = \frac{p!S_{[m]}}{m!(p-m)!} \approx \frac{\optlam^{m}}{m!},\label{eq:Sk_approx1}
\end{align}
where $\optlam$ is defined in (\ref{eq:lambda}). Then combining (\ref{eq:Pk_def}) and
(\ref{eq:Sk_approx1}), we have
\begin{align}
\P\left(\tneb=k\right) & =\sum_{m=0}^{p-k}\binom{k+m}{k}(-1)^{m}S_{k+m}\nonumber \\
 & \approx\sum_{m=0}^{p-k}\frac{(k+m)!}{m!k!}(-1)^{m}\frac{\lambda_{p}^{k+m}}{(k+m)!}\nonumber \\
 & \approx\frac{\lambda_{p}^{k}}{k!}e^{-\lambda_{p}},\label{eq:Pm_approx1}
\end{align}
which implies that the PMF of $\tneb$ is approximately Poisson with
rate $\lambda_{p}$.

We now quantitatively analyze the error of approximation in (\ref{eq:Pm_approx1}).
First, we approximate the right-hand side of (\ref{eq:Pk_def}) by a truncated sum: $\sum_{m=k}^{L}\binom{m}{k}(-1)^{m-k}S_{m}$, with $L\leq p$.
The reason for this operation is that $S_{[m]}\approx \Phi^{m}\left(-\tfrac{1}{\opttau}\right)$ may not be accurate for large $m$, since we only have approximate finite event independence. We then need to control the error caused by the truncation. Accordingly,
we can apply Bonferroni's inequality \cite[p.110]{feller1968introduction}, stated as follows.
Under the same setting as (\ref{eq:Pk_def}), for $k+1\leq L\leq p$, we have
\begin{enumerate}
\item If $L-k$ is odd,
\begin{equation}
\sum_{m=k}^{L}\binom{m}{k}(-1)^{m-k}S_{m}\leq P_{k}\leq\sum_{m=k}^{L-1}\binom{m}{k}(-1)^{m-k}S_{m}.\label{eq:Bonferroni_1}
\end{equation}
\item If $L-k$ is even,
\begin{equation}
\sum_{m=k}^{L-1}\binom{m}{k}(-1)^{m-k}S_{m}\leq P_{k}\leq\sum_{m=k}^{L}\binom{m}{k}(-1)^{m-k}S_{m}.\label{eq:Bonferroni_2}
\end{equation}
\end{enumerate}
Therefore, we need to choose a reasonably large $L$ to attain a good trade-off between the approximation error of (\ref{eq:Sk_approx1}) and the truncation error of (\ref{eq:Bonferroni_1}) and (\ref{eq:Bonferroni_2}), such that they are both properly bounded.
Our proof of Proposition \ref{prop:Poisson_convergence} follows this idea. The details can be found in Appendix \ref{appen:tneb_proof}.

\section{Conclusion}
\label{sec:Conclusions} 
In this paper, we have presented an exact performance characterization of the box-relaxation decoder in high dimensions. We show that, under certain scalings of the sampling ratio and the noise variance, the number of incorrectly-decoded bits has a limiting Poisson distribution. In addition, a phase transition from nonperfect to perfect recovery takes place at a well-defined critical threshold. Numerical simulations show that the actual performance of the algorithm is well captured by our theoretical predictions. Finally, it is worth mentioning that, although we have assumed that the sensing matrix has i.i.d. normal entries, the results on the limiting Poisson law should hold under more general matrix ensembles. We leave this as an interesting line of work for future investigation.

\appendix

\subsection{Properties of $\mathcal{G}_{p}(\vs)$ and $g_{p}(v)$}
\label{appendix:Gpsgpv}
\begin{lem}
\label{lem:Gp_v}
For any $\dmtx$ and $\vy$, it holds that:
\begin{enumerate}
\item $\mathcal{G}_{p}(\vs)$ is convex and differentiable in $\R^{n}$, with
\begin{align}
\nabla\mathcal{G}_{p}(\vs) & =\vu_{\vs}^{*}-\vu^{*},\label{eq:gradient_Gp}
\end{align}
where $\vu_{\vs}^{*}\bydef\argmax{\vu}\,\vs^{\T}\vu-\Lcav(\vu).$
\item $\nabla\mathcal{G}_{p}(\vs)$ is 1-Lipschitz continuous, i.e., $\forall \vr,\vs \in \R^{n}$
\begin{equation}
\|\nabla\mathcal{G}_{p}(\vr)-\nabla\mathcal{G}_{p}(\vs)\|\leq\|\vr-\vs\|\label{eq:Gp_Lipschitz}
\end{equation}
or equivalently,
\begin{equation}
\|\vu_{\vr}^{*}-\vu_{\vs}^{*}\|\leq\|\vr-\vs\|.\label{eq:ux_Lipschitz}
\end{equation}
\item
$g_{p}(v)$ is convex and differentiable with
\begin{align}
\left|g_{p}'(v)\right| & \leq2\|\vacav\|^{2},\label{eq:gp_derivative_upperbd}
\end{align}
\end{enumerate}
\end{lem}
\begin{IEEEproof}
Let $\Lcav^{*}(\vs) \bydef \max_{\vu}\vs^{\T}\vu-\Lcav(\vu)$, which is the conjugate function of $\Lcav(\vu)$. We know $\nabla\mathcal{G}_{p}(\vs) = \nabla\Lcav^{*}(\vs) - \vu^{*}$. Since $\Lcav(\vu)$ is closed and 1-strongly convex, $\Lcav^{*}(\vs)$ is convex and differentiable with $\nabla\Lcav^{*}(\vs) = \vu_{\vs}^{*}$ and $\nabla\Lcav^{*}(\vs)$ is 1-Lipschitz continuous \cite[Chapter X]{hiriartconvex}. Therefore, from (\ref{eq:Gpx}) we know $\mathcal{G}_{p}(\vs)$ is convex. Since $\nabla\mathcal{G}_{p}(\vr) - \nabla\mathcal{G}_{p}(\vs) = \nabla\Lcav^{*}(\vr) - \nabla\Lcav^{*}(\vs)$, we get (\ref{eq:gradient_Gp}) and (\ref{eq:Gp_Lipschitz}).

Since $g_{p}(v)  = \mathcal{G}_{p}(\vacav v)$, $g_{p}(v)$ is also convex and differentiable with $g_p'(v)=\vacav^{\T}\nabla\mathcal{G}_{p}(\vacav v)$. From \eqref{eq:gradient_Gp} and \eqref{eq:Gp_Lipschitz}, we know $\|\nabla\mathcal{G}_{p}(\vacav v)\| \leq \|\vacav\|v$. Therefore,  \eqref{eq:gp_derivative_upperbd} follows from Cauchy-Schwartz inequality and the fact that $|v|\leq 2$. 
\end{IEEEproof}

\subsection{Properties of the Optimization Problem (\ref{eq:tau})}
\label{appen:1Dopt}
In this section, we collect some useful properties of the one-dimensional optimization (\ref{eq:tau}), which was first studied in \cite{thrampoulidis2018symbol}. For our purpose, we consider a slightly more general setting:
\begin{align}
\label{eq:fpstar_t}
f_p(t) &= \min_{\tau>0} F_p(\tau; t, \sampratio) \nonumber \\
&= \min_{\tau>0}\frac{\tau}{2}\left(\sampratio-\frac{1}{2}\right)+\frac{t}{2\tau}+\frac{\tau}{2}\int_{\frac{2}{\tau}}^{\infty}\left(x-\frac{2}{\tau}\right)^{2}\Phi(dx),
\end{align}
where $t > 0$ is a parameter. Note that (\ref{eq:tau}) and the inline optimization of (\ref{eq:Qpstar_theta}) are the cases where $t=\nsdp^2$ and $t=(1+\field)^2\nsdp^2$, respectively. Also we define the squared loss function: $R_p(t) \bydef \frac{f_p^2(t)}{2}$ and evidently, $R_p\big[(1+\field)^2\nsdp^2\big] = \optQ(\field)$, where $\optQ(\field)$ is defined in (\ref{eq:Qpstar_theta}).
\subsubsection{Uniqueness of Optimal Solution}
Let $\tau(t)$ be the minimizer of (\ref{eq:fpstar_t}), which is the solution of stationary equation:
\begin{equation}
h(\tau) \bydef \sampratio-\frac{1}{2}+\int_{\frac{2}{\tau}}^{\infty}\left(x^{2}-\frac{4}{\tau^{2}}\right)\Phi(dx)-\frac{t}{\tau^{2}}=0.\label{eq:tau_t_equation}
\end{equation}
By direct differentiation of $h(\tau)$ above, we can show $h'(\tau) = \int_{\frac{2}{\tau}}^{\infty}\frac{8}{\tau^3}\Phi(dx)+\frac{2t}{\tau^3}>0$, so it is a strictly increasing function. Also $\lim_{\tau\to 0}h(\tau) = -\infty$ and $\lim_{\tau\to\infty}h(\tau) = \sampratio>0$. This also establishes that the strict convexity of $f_p(t)$. Therefore, $\tau(t)$ is unique for any $t>0$. Besides, we can directly check that $\tau(t)$ is differentiable with
\begin{equation}\label{eq:tau_deri}
\tau'(t)=\frac{\tau(t)}{8\int_{2/\tau(t)}^{\infty}\Phi(dx)+2t}>0,
\end{equation}
so $\tau(t)$ is strictly increasing.
\subsubsection{Upper and Lower Bounds of $\tau(t)$}
Since $h\Big(\sqrt{\tfrac{t}{\sampratio}}\Big) < -\frac{1}{2}+\int_{\frac{2}{\tau}}^{\infty}x^{2}\Phi(dx)<0$, by $h(0^+)<0$, $h(\infty)>0$ and uniqueness of $\tau(t)$, we have $\tau(t)\geq\sqrt{\tfrac{t}{\sampratio}}$. Similarly, we can get $\tau(t)\leq\min\{\sqrt{\tfrac{t}{\sampratio-1/2}},\sqrt{\tfrac{4+t}{\sampratio}}\}$ and $\tau(t)\geq\sqrt{\tfrac{t}{\sampratio-1/2+v_{p}}}
$, where $v_{p} = \int_{b_p}^{\infty}x^{2}\Phi(dx)$, with $b_p = 2\sqrt{\tfrac{\sampratio-1/2}{t}}$ and evidently, $v_{p}<1/2$. Therefore, $\tau(t)$ can be bounded as:
\begin{equation}
\label{eq:tau_t_ublb}
\sqrt{\tfrac{t}{\sampratio-1/2+v_{p}}} \leq \tau(t) \leq \min\Big\{\sqrt{\tfrac{t}{\sampratio-1/2}},\sqrt{\tfrac{4+t}{\sampratio}}\Big\}.
\end{equation}
\subsubsection{Properties of $f_p(t)$}
From (\ref{eq:fpstar_t}) we get $f_p(t)\geq 0$, $f_p'(t)=\frac{1}{2\tau(t)}>0$ and $f_p''(t)=-\frac{\tau'(t)}{2\tau^2(t)}<0$, so $f_p(t)$ is nonnegative, strictly increasing and concave. On the other hand, letting $\tau=\sqrt{\tfrac{t}{\sampratio}}$ in (\ref{eq:fpstar_t}) we can get $f_p(t)\leq C \big(\frac{\sqrt{t\sampratio}}{2}+1\big)$, where $C$ is some constant.
\subsubsection{Properties of $R_p(t)$}
By the chain rule, $R_p'(t)=\frac{f_p(t)}{2\tau(t)}$ and $R_p''(t) = \tfrac{\int_{2/\tau(t)}^{\infty}x\Phi(dx)}{\tau(t)(8\int_{2/\tau(t)}^{\infty}\Phi(dx)+2t)}$. Therefore, $R_p(t)$ is strictly increasing and convex. From (\ref{eq:tau_t_equation}), we can show $R_p'(t)$ is bounded:
\begin{equation}\label{eq:Rpderi_bd}
R_p'(t) = \frac{1}{2}\left[\sampratio-\tfrac{1}{2}+\int_{\frac{2}{\tau(t)}}^{\infty}x^{2}-\tfrac{2x}{\tau(t)}\Phi(dx)\right]\leq\frac{\sampratio}{2}.
\end{equation}
On the other hand, $R_p''(t)$ satisfies: $R_p''(t)\leq \frac{\varphi(-2/\tau(t))}{2\tau(t)t}$, where $\varphi(x)$ is the PDF of standard Gaussian. Then using \eqref{eq:tau_t_ublb} and Assumption \ref{asmp:A3}, we know there exists $C>0$, s.t., for $t>0$,
\begin{align}\label{eq:Rp_2ndDeri_bd}
R_p''(t) \leq \sqrt{\tfrac{\sampratio}{8\pi}}e^{-\tfrac{2(\sampratio-1/2)}{t}}t^{-\tfrac{3}{2}} \leq C.
\end{align}
%
\subsection{Proof of Proposition \ref{prop:uniform_convergence}}
\label{appen:uniformconv}
We first prove the pointwise convergence of $g_{p}(v)$ to $\E g_{p}(v)$: there exists $c>0$, s.t., for any $v\in[0,2]$ and $\veps>0$,
\begin{equation}
\P\left(\left|g_{p}(v)-\E g_{p}(v)\right|>\varepsilon\right)\leq ce^{-c^{-1}p\min\left\{ \tfrac{\veps^2}{\sampratio},\veps \right\} }.\label{eq:concentration_Gp_pointwise}
\end{equation}

Recall that $g_{p}(v)=\mathcal{G}_{p}\left( {\vacav}{v} \right)$, so it is equivalent to prove $|\mathcal{{G}}_{p}({\vacav}v)-\E\mathcal{{G}}_{p}({\vacav}v)|\to 0$. We first control the moment generating function of $\mathcal{{G}}_{p}({\vacav}v)-\E\mathcal{{G}}_{p}({\vacav}v)$. Let $\vb$ be an i.i.d. copy of $\va$. For all $|\lambda|\leq\frac{p}{2\sqrt{2}\pi}$, we can apply Theorem 2.2 of \cite[p.176]{pisier1986probabilistic} to get
\begin{align*}
\E[\exp\lambda(\mathcal{{G}}_{p}({\vacav}v)-\E\mathcal{{G}}_{p}({\vacav}v))] &\leq \E e^{ \frac{\pi\lambda}{2}(\vb v)^{\T}\nabla\mathcal{{G}}_{p}({\vacav}v) }\\
&\tleq{\text{(a)}}\E_{\dmtx,\noise}\E_{\vacav}e^{\frac{2\lambda^{2}\pi^{2}}{p}\|\vacav\|^{2}}\\
& =\exp\left[-\frac{n}{2}\log\left(1-\frac{4\lambda^{2}\pi^{2}}{p^{2}}\right)\right]\\
& \tleq{\text{(b)}}\exp\left(\frac{4\sampratio\lambda^{2}\pi^{2}}{p}\right).
\end{align*}
In step (a), we take expectation over $\vb$ and use $|v|\leq 2$ and $\|\nabla\mathcal{{G}}_{p}({\vacav}v)\| \leq {2\|{\vacav}\|}$, as implied by (\ref{eq:Gp_Lipschitz}); In step (b), we use the inequality $\log(1+x)\geq\frac{x}{1+x}$,
for $x>-1$ and the condition that $|\lambda|\leq\frac{p}{2\sqrt{2}\pi}$.
As a result, for any $\veps \geq 0$ and $\lambda\in\left[0,\frac{p}{2\sqrt{2}\pi}\right]$,
\begin{align}
\label{eq:Chernoff1}
\P\left( g_{p}(v)-\E g_{p}(v) > \varepsilon \right) & \leq e^{-\lambda\varepsilon+\frac{4\sampratio\lambda^{2}\pi^{2}}{p}}.
\end{align}
After minimizing the exponent on the RHS of (\ref{eq:Chernoff1})
over $\lambda\in\left[0,\frac{p}{2\sqrt{2}\pi}\right]$,
we can get for any $\veps\in[0,\sqrt{8}\pi\sampratio]$, $\P\left(g_{p}(v)-\E g_{p}(v)>\varepsilon\right)\leq e^{-\tfrac{p\varepsilon^{2}}{16\sampratio\pi^{2}}}$; for any $\veps > \sqrt{8}\pi\sampratio$, $\P\left(g_{p}(v)-\E g_{p}(v)>\varepsilon\right)\leq e^{-\tfrac{p\varepsilon}{4\sqrt{2}\pi}}$. The other direction also holds by the same reasoning. Thus,
\begin{align}
\label{eq:gpv_concentrate}
\P\left(\left|g_{p}(v)-\E g_{p}(v)\right|>\varepsilon\right)&\leq 2 e^{-\tfrac{p}{16\pi^2}\min\{\tfrac{\veps^2}{\sampratio }, \veps\}}.
\end{align}

To show uniform convergence (\ref{eq:concentration_gp_uniform}), it suffices to prove the Lipschitz continuity of $g_{p}(v)$ and $\E g_{p}(v)$. From Lemma 1 of \cite{laurent2000adaptive}, we have for all $x>0$, $\P\Big(\|\vacav\|^2\geq \sampratio +\tfrac{2\sqrt{\sampratio x}}{\sqrt{p}}+\tfrac{2x}{p}\Big)\leq \exp(-x)$.
Let $x=n(\sqrt{y+1}-1)^2$, we have for $y\geq 2$, $\P\left(  \tfrac{\|\va\|^2}{\sampratio}-1\geq y\right)\leq\exp\left(-\tfrac{ny}{4}\right)$.
Therefore, by taking $y=K/\sampratio$, we get for any $K\geq2\sampratio$,
\begin{align}
\label{eq:K_concentrate}
\P(\|\vacav\|^{2}>K) \leq \P(\|\vacav\|^{2}-\sampratio>K/2) \leq \exp(-\tfrac{pK}{4}).
\end{align}
Combining it with (\ref{eq:gp_derivative_upperbd}), we know for $K\geq2 \sampratio$, $g_{p}(v)$ is $2K$-Lipschitz
with probability greater than $1-\exp(-\tfrac{pK}{4})$.
From (\ref{eq:gp_derivative_upperbd}), we can also get $\left|\frac{d\E g_{p}(v)}{dv}\right|\leq 2 \sampratio$, so $\E g_{p}(v)$
is $2 \sampratio$-Lipschitz continuous over $v\in[0,2]$. Combining the Lipschitz continuity
of $g_{p}(v)$ and $\E g_{p}(v)$ with (\ref{eq:concentration_Gp_pointwise}),
we can obtain (\ref{eq:concentration_gp_uniform}) by a standard epsilon-net argument as follows. We need to consider different values of $\veps$:
\begin{enumerate}
  \item If $\veps \geq \sampratio$, we construct an epsilon-net of $[0,2]$ formed by the following points: $v_k = \frac{k}{4}$, $k=1,2,\ldots,8$. For any $v\in[0,2]$, denote $v^{*}$ as the closest point to $v$ in the above epsilon-net. By construction, $|v-v^{*}|\leq \frac{1}{8}$. If $g_{p}(v)$ is $2K$-Lipschitz, then for any $v\in[0,2]$,
      \begin{align}
      |g_{p}(v)-\E g_{p}(v)| &\leq |g_{p}(v)-g_{p}(v^{*})| + |g_{p}(v^{*})-\E g_{p}(v^{*})| + |\E g_{p}(v^{*})-\E g_{p}(v)|\nonumber\\
      &\leq \frac{K}{4} + |g_{p}(v^{*})-\E g_{p}(v^{*})| + \frac{\veps}{2},
      \end{align}
      where we have used the Lipschitz continuity of $g_{p}(v)$ and $\E g_{p}(v)$, as well as $\veps \geq \sampratio$. Then $\sup_{v\in[0,2]}|g_{p}(v)-\E g_{p}(v)|\geq 2\veps$, only if at least one of following holds: (i) $K\geq 2\veps\geq 2\sampratio$, (ii) there exists a $k\in\{1,2,\ldots,8\}$, s.t., $|g_{p}(v_{k})-\E g_{p}(v_{k})| \geq \veps$. Combining (\ref{eq:gpv_concentrate}) and (\ref{eq:K_concentrate}) and applying the union bound, we get for $\veps\geq \sampratio$,
      \begin{equation}
      \label{eq:gpv_concentrate1}
       \P(\sup_{v\in[0,2]}|g_{p}(v)-\E g_{p}(v)|\geq 2\veps)\leq 18 e^{-\tfrac{p\veps}{16\pi^2} }.
      \end{equation}
  \item If $\veps< \sampratio$, we construct an epsilon-net of $[0,2]$ formed by the following points: $v_k = 2k/\lceil\frac{8\sampratio}{\veps}\rceil$, $k=1,2,\ldots,\lceil\frac{8\sampratio}{\veps}\rceil$. In this case, for any $v\in[0,2]$, we have $|v-v^{*}|\leq \frac{\veps}{8\sampratio}$. Then similar as previous argument, we have $\sup_{v\in[0,2]}|g_{p}(v)-\E g_{p}(v)|\geq 2\veps$, only if at least one of following holds: (i) $g_{p}(v)$ is not $4\sampratio$-Lipschitz, (ii) there exists a $k \leq \lceil\frac{8\sampratio}{\veps}\rceil$, s.t., $|g_{p}(v_{k})-\E g_{p}(v_{k})| \geq \veps$. Combining (\ref{eq:gpv_concentrate}) and (\ref{eq:K_concentrate}) and applying the union bound, we get:
      \begin{equation}
      \label{eq:gpv_concentrate2}
       \P(\sup_{v\in[0,2]}|g_{p}(v)-\E g_{p}(v)|\geq 2\veps)\leq \frac{16\sampratio}{\veps} e^{-\tfrac{p\veps^2}{16\pi^2\sampratio} }.
      \end{equation}
\end{enumerate}
Combining (\ref{eq:gpv_concentrate1}) and (\ref{eq:gpv_concentrate2}),
together with symmetry and the union bound, we directly get (\ref{eq:concentration_gp_uniform}). 
\subsection{Proof of Lemma \ref{lem:Qp_theta_concentration}}
\label{appendix:CGMTproof}
The proof follows the CGMT framework \cite{thrampoulidis2016ber,thrampoulidis2018symbol}. The optimization in (\ref{eq:Qp_theta_def}) is equivalent to
\begin{align}
\losspo p(\field)
= & {p^{-\tfrac{3}{2}}}\min_{\vx\in[-1,1]^{p}}\max_{\vu}\vu^{\T}\begin{bmatrix}\sqrt{p}\dmtx & -\widetilde{\vw}\end{bmatrix}\begin{bmatrix}\vx-\sgl\\
\sqrt{p(\field + \nsdp^2)}
\end{bmatrix}-\frac{\sqrt{p}\|\vu\|^{2}}{2},\label{eq:CGMT_PO}
\end{align}
where $\widetilde{\vw}\sim\mathcal{N}(\boldsymbol{0},\mI_n)$. The corresponding auxiliary problem (AO) of (\ref{eq:CGMT_PO}) is
\begin{align}
Q_{\text{AO},p}(\field)&=\min_{\vx\in[-1,1]^{p}}\max_{\vu}-\sqrt{\tfrac{\|\vx-\sgl\|^{2}}{p}+\field + \nsdp^2}\,\tfrac{\vg^{\T}\vu}{p} +\tfrac{\|\vu\|}{\sqrt{p}}\left[\tfrac{\vh^{\T}(\vx-\sgl)}{p}+\tfrac{h_{0}\sqrt{\field + \nsdp^2}}{\sqrt{p}}\right]-\tfrac{\|\vu\|^{2}}{2p}\nonumber\\
&=  \frac{1}{2}\left(\min_{\vx\in[-1,1]^{p}}\sqrt{\tfrac{\|\vx-\sgl\|^{2}}{p}+\field + \nsdp^2}\,\tfrac{\|\vg\|}{\sqrt{p}}+\tfrac{\vh^{\T}(\vx-\sgl)}{p}+\tfrac{h_{0}\sqrt{\field + \nsdp^2}}{\sqrt{p}}\right)_{+}^{2}.\label{eq:CGMT_AO_1}
\end{align}
where $(x)_{+}\bydef \max\{x,0\}$, $\vg\sim\mathcal{N}(\boldsymbol{0},\mI_{n}),\vh\sim\mathcal{N}(\boldsymbol{0},\mI_{p})$,
$h_{0}\sim\mathcal{N}(0,1)$ and they are mutually independent.

Now we analyze the inline optimization problem of (\ref{eq:CGMT_AO_1}),
which can be simplified as:
\begin{align}
\phi(\field,\vg,\vh)= & \min_{\vx\in[-1,1]^{p}}\inf_{\tau > 0}\sqrt{\sampratio}\left[\tfrac{\tau}{2}+\tfrac{\|\vx-\sgl\|^{2}}{2\tau p}+\tfrac{\field+\nsdp^2}{2\tau}\right]\tfrac{\|\vg\|}{\sqrt{n}}+\tfrac{\vh^{\T}(\vx-\sgl)}{p}+\tfrac{h_{0}\sqrt{\field+\nsdp^2}}{\sqrt{p}}\label{eq:CGMT_AO_2}\\
= & \inf_{\tau> 0}\underbrace{\left[\tfrac{\tau\sampratio}{2}+\tfrac{\field+\nsdp^2}{2\tau}\right]\tfrac{\|\vg\|}{\sqrt{n}}+\tfrac{1}{p}\sum_{i=1}^{p}v\left(h_{i};\tau,\vg\right)+\tfrac{h_{0}\sqrt{\field+\nsdp^2}}{\sqrt{p}}}_{F(\tau;\field,\vg,\vh)},\label{eq:CGMT_AO_3}
\end{align}
where in (\ref{eq:CGMT_AO_3}) we make a change of variable: $\tfrac{\tau}{\sqrt{\sampratio}}\to\tau$
and the parametric function $v\left(h;\tau,\vg\right)$
is defined as:
\begin{equation}
\label{eq:v_h}
v\left(h;\tau,\vg\right)\bydef\begin{cases}
0 & h\geq0,\\
-\frac{\tau\sqrt{n}}{2\|\vg\|}h^{2} & h\in[\frac{-2\|\vg\|}{\tau\sqrt{n}},0),\\
2\left(\frac{\|\vg\|}{\tau\sqrt{n}}+h\right) & h<\frac{-2\|\vg\|}{\tau\sqrt{n}}.
\end{cases}
\end{equation}
Denote $\tauAO(\field)$ as the optimal solution in (\ref{eq:CGMT_AO_3}). From (\ref{eq:CGMT_AO_2}) and the fact that we did a change of variable in (\ref{eq:CGMT_AO_3}), it can be seen $\tauAO(\field) =\sqrt{\tfrac{\|\vx^{*}-\sgl\|^{2}}{p\sampratio}+\tfrac{\field+\nsdp^2}{\sampratio}}$.
Therefore, for $\field\in[0,1]$, $\tauAO(\field)\in\Omega(\nsdp,\sampratio)$
, where $\Omega(\nsdp,\sampratio)\bydef\left[\tfrac{\nsdp}{\sqrt{\sampratio}},\tfrac{\sqrt{5+\sigma_{p}^{2}}}{\sqrt{\sampratio}}\right]$. Note that this is consistent with (\ref{eq:tau_t_ublb}).

We now show objective function $F(\tau;\field,\vg,\vh)$ in (\ref{eq:CGMT_AO_3}) converges to
$F(\tau;\field) \bydef F_p(\tau; \field+\nsdp^2, \sampratio)$
with high probability over $\tau \in\Omega(\nsdp,\sampratio)$. The first and third term in RHS of (\ref{eq:CGMT_AO_3}) is relatively easy to deal with. By the concentration of $\tfrac{\|\vg\|}{\sqrt{n}}$  (e.g. \cite[p.44]{vershynin2018high}) and $\tfrac{h_0}{\sqrt{p}}$, there exists $c>0$, s.t., for any $\veps>0$ and $\tau \in\Omega(\nsdp,\sampratio)$,
\begin{align}\label{eq:gnorm_conv}
\P\left(\left(\tfrac{\tau\sampratio}{2}+\tfrac{\field+\nsdp^2}{2\tau}\right) \left|\tfrac{\|\vg\|}{\sqrt{n}} - 1\right| > \sqrt{\sampratio}\veps\right) \leq c\exp(-n\veps^2/c)
\end{align}
and
\begin{align}
\label{eq:h0_conv}
\P\left(\left|\tfrac{h_{0}\sqrt{\field+\nsdp^2}}{\sqrt{p}}\right|>\veps\right)\leq c\exp(-p\veps^2/c).
\end{align}
Here in \eqref{eq:gnorm_conv}, we have used the fact that for $\tau\in\Omega(\nsdp,\sampratio)$, $\tfrac{\tau\sampratio}{2}+\tfrac{\field+\nsdp^2}{2\tau}\leq C\sqrt{\sampratio}$, where $C$ is some constant. For the second term, define the following function: $V(\vh;\tau,\vg) \bydef \tfrac{\sum_{i=1}^{p}v\left(h_{i};\tau, \vg \right)}{p}$, where $v\left(h;\tau,\vg\right)$ is given in (\ref{eq:v_h}).  We now show there exists $c>0$, s.t., for any $\veps\geq 0$,
\begin{equation}
\label{eq:Vh_conv1}
\P( |V(\vh;\tau,\vg)- f(\tau)| > \veps) \leq c \exp(-{p\veps^2}/{c}),
\end{equation}
where 
\begin{equation}
\label{eq:tauvg}
f(t) \bydef {-\tfrac{t}{4} + \tfrac{t}{2}\int_{{2}/{t}}^{\infty}\left(x-\tfrac{2}{t}\right)^{2}\Phi(dx)}.
\end{equation}
First, note that for any fixed $\vg$, $v\left(h;\tau,\vg\right)$ is 2-Lipschitz continuous, so $V(\vh;\tau,\vg)$ is $\tfrac{2}{\sqrt{p}}$-Lipschitz continuous w.r.t. $\vh$. Also we can verify that $\E_h v(h;\tau,\vg) = f(\tau_{\vg})$, with $\tau_{\vg}\bydef \frac{\tau\sqrt{n}}{\|\vg\|}$. Then using Theorem 2.1 in \cite[p.176]{pisier1986probabilistic}, we have for any $\vg$ and $\veps>0$,
\begin{equation}
\label{eq:Vh_conv}
\P( |V(\vh;\tau,\vg)-f(\tau_{\vg}) | > \veps) \leq 2 \exp(-\tfrac{p\veps^2}{2\pi^2}).
\end{equation}
It can be checked that $f(t)$ in (\ref{eq:tauvg}) satisfies $f(t)\in [-1, 0]$ for any $t>0$. Combining this with (\ref{eq:tauvg}) and (\ref{eq:Vh_conv}), we know \eqref{eq:Vh_conv1} holds for $\veps>\tfrac{1}{2}$.
On the other hand, by a direct differentiation, we have $f'(t)=-\tfrac{1}{4}+\tfrac{1}{2}\int_{{2}/{t}}^{\infty} \big(x^2 - \tfrac{4}{t^2}\big)\Phi(dx)$. It is not hard to verify $|f'(t)|\leq 1/4$, for all $t>0$. Therefore, for any $\veps \in (0,1/2)$, on the event $E_{\veps} = \left\{\left|\tfrac{\|\vg\|}{\sqrt{n}}-1\right|<\veps\right\}$, which happens with probability $\P(E_{\veps})\geq 1 - c e^{-n\veps^2/c}$, there exists $c>0$, s.t., $|\tau_{\vg}-\tau|\leq c\veps$. As a result, there exists $c>0$, s.t., for $\veps\in(0,1/2)$, $\P(|f(\tau_{\vg})-f(\tau)|>\veps)\leq c e^{-n\veps^2/c}$. This together with (\ref{eq:Vh_conv}) implies there exists $c>0$, s.t., for $\veps\in(0,1/2)$, inequality (\ref{eq:Vh_conv1}) still holds.

Combining (\ref{eq:gnorm_conv}) and (\ref{eq:Vh_conv1}), we get that there exists $c>0$, s.t., for any $\veps>0$, $\tau\in\Omega(\nsdp,\sampratio)$ and $\field\in[0,1]$,
\begin{equation}
\P\left(|F(\tau;\field,\vg,\vh)-F(\tau;\field)|>\veps\right)\leq ce^{-p\veps^{2}/c}.\label{eq:F_tau_convergence}
\end{equation}
On the other hand, it can be verified from definition that there exists $C>0$, s.t., $F(\tau;\field,\vg,\vh)$ and $F(\tau;\field)$ are both $C{\sampratio}$-Lipschitz over $\tau\in\Omega(\nsdp,\sampratio)$. Then by a similar epsilon-net argument as in the proof of Proposition \ref{prop:uniform_convergence}, we can get:
\begin{equation}
\P\big(\sup_{\tau\in\Omega(\nsdp,\sampratio)}|F(\tau;\field,\vg,\vh)-F(\tau;\field)|>\veps\big)\leq \tfrac{c\sqrt{\sampratio}}{\veps}e^{-p\veps^{2}/c}.\label{eq:F_tau_theta_uniform_convergence}
\end{equation}
Since $\phi(\field,\vg,\vh)=\min_{\tau\in\Omega(\nsdp,\alpha_{p})}F(\tau;\field,\vg,\vh)$
and $\sqrt{2Q_{p}^{*}(\field)}=\min_{\tau\in\Omega(\nsdp,\alpha_{p})}F(\tau;\field)$,
from (\ref{eq:F_tau_theta_uniform_convergence}) we know there exists
$c>0$, s.t., for any $\veps > 0$,
\begin{equation}
\P\left(|\phi(\field,\vg,\vh)-\sqrt{2Q_{p}^{*}(\field)}|>\veps\right)\leq\tfrac{c\sqrt{\sampratio}}{\veps}e^{-p\veps^{2}/c}.\label{eq:sqrt_Qpstar_convergence}
\end{equation}
Since $\sqrt{2Q_{\text{AO},p}(\field)}=\max\{\phi(\field,\vg,\vh),0\}$,
from (\ref{eq:sqrt_Qpstar_convergence}) we have
\begin{equation}
\label{eq:sqrt_Qpstar_convergence2}
\P\left(|\sqrt{2Q_{\text{AO},p}(\field)}-\sqrt{2Q_{p}^{*}(\field)}|>\veps\right) \leq \tfrac{c\sqrt{\sampratio}}{\veps}e^{-p\veps^{2}/c}.
\end{equation}
Taking into account the fact $Q_{p}^{*}(\field)\leq C{\sampratio}$ (as shown in Appendix \ref{appen:1Dopt}), we
can further obtain the following Bernstein's type inequality: there exists $ c>0$,
s.t., for any $\veps>0$ and $\field\in[0, 1]$,
\begin{align}
\P\left(|Q_{\text{AO},p}(\field)-Q_{p}^{*}(\field)|>\veps\right)\leq &
\frac{ce^{-p\min\{\tfrac{\veps^2}{\sampratio }, \veps\}/c}}{\min\{\tfrac{\veps}{\sampratio},\sqrt{\tfrac{\veps}{\sampratio}}\}}.\label{eq:QAOp_theta_concentration_1}
\end{align}
Then by CGMT (e.g., \cite[Corollary 5.1]{miolane2018distribution}), \eqref{eq:QAOp_theta_concentration_1} implies that there exists
$c>0$, s.t.,
\begin{equation}
\P\left(|\losspo p(\field)-Q_{p}^{*}(\field)|>\veps\right)\leq
\frac{ce^{-p\min\{\tfrac{\veps^2}{\sampratio }, \veps\}/c}}{\min\{\tfrac{\veps}{\sampratio},\sqrt{\tfrac{\veps}{\sampratio}}\}}.\label{eq:Qp_theta_concentration_1}
\end{equation}
Finally, from (\ref{eq:Qp_theta_concentration_1}) we know there exists $c>0$, s.t., for any
$\eta>0$ and $\field\in[0,1]$,
\begin{align}
\E|\losspo p(\field)-Q_{p}^{*}(\field)| & =\int_{0}^{\infty}\P\left(|\losspo p(\field)-Q_{p}^{*}(\field)|\geq t\right)dt\nonumber \\
 & \leq\sqrt{\sampratio}\eta+\int_{\sqrt{\sampratio}\eta}^{\infty}\frac{c\sampratio}{t}e^{-\tfrac{pt^{2}}{c{\sampratio}}}dt
 +\int_{\sqrt{\sampratio}\eta}^{\infty}c \sqrt{\frac{\sampratio}{t}} e^{-\tfrac{pt}{c}} dt\\
 & \leq \sqrt{\sampratio}\eta+\frac{c^{2}\sampratio}{p}\left( \frac{e^{-p\eta^{2}/c}}{\eta^{2}} + \frac{e^{-\sqrt{\sampratio}p\eta/c}}{\sqrt{\eta}} \right).\label{eq:E_absdiff_Qp_Qpstar}
\end{align}
Then for $\expnt>2$, letting $\eta=p^{-1/\expnt}$ in (\ref{eq:E_absdiff_Qp_Qpstar}) and taking into account Assumption \ref{asmp:A3},
we can get $\E|\losspo p(\field)-Q_{p}^{*}(\field)|\leq cp^{-1/\expnt}$
for some $c>0$ and all the sufficiently large $p$. As a result,
\[
|\averlosspo p(\field)-Q_{p}^{*}(\field)|\leq\E|\losspo p(\field)-Q_{p}^{*}(\field)|\leq cp^{-1/\expnt}.
\]
Since the constant $c$ above does not depend on $\field$, we get
\eqref{eq:EQp_theta_concentration}. 
\subsection{Approximate $k$-wise Independence }
\label{appen:ApproximateIndependence}
\subsubsection{$\big\{ \va_{i}^{\T}\vu_{\backslash i}^{*}\big\} _{i\in[k]} \tapprox{d} \big\{ \va_{i}^{\T}\vu_{\backslash[k]}^{*}\big\} _{i\in[k]}$}
We first prove that the joint distribution of $\big\{ \va_{i}^{\T}\vu_{\backslash i}^{*}\big\} _{i\in[k]}$
is close to $\big\{ \va_{i}^{\T}\vu_{\backslash[k]}^{*}\big\} _{i\in[k]}$. To prove this, we can show $\va_{i}^{\T}\vu_{\backslash i}^{*}\approx \va_{i}^{\T}\vu_{\backslash [k]}^{*}$, for any $i\in[k]$ and use the fact that $\va_{i}^{\T}\vu_{\backslash i}^{*}$ and $\va_{i}^{\T}\vu_{\backslash [k]}^{*}$ are in the same probability space.
\begin{lem}
\label{lem:a1t_deltaUn1_concentrate}There exists $c>0$, s.t., for any $\veps>0$
and $i=1,2,\ldots,p-1$,
\begin{equation}
\P\left(\left|\va_{i}^{\T}(\vu_{\backslash[i]}^{*}-\vu_{\backslash[i+1]}^{*})\right|>\sqrt{\sampratio}\varepsilon\right)\leq ce^{-c^{-1} p\min\left\{ \veps^{2},\veps\right\}}.\label{eq:a1t_deltaUn1_concentrate}
\end{equation}
\end{lem}
\begin{IEEEproof}
To lighten notation, define $\Delta_{[i]}\bydef\vu_{\backslash[i]}^{*}-\vu_{\backslash[i+1]}^{*}$.
Denote the objective function in \eqref{eq:leave_k_out_Lu} as $L_{\backslash [i]}(\vu)$, (with $k$ replaced by $i$ here). By strong convexity of $L_{\backslash [i]}(\vu)$, we have
\begin{align}
\label{eq:Lcavi_strongconv1}
L_{\backslash [i]}(\vu_{\backslash [i+1]}^{*}) - L_{\backslash [i]}(\vu_{\backslash [i]}^{*}) \geq \tfrac{1}{2}\|\Delta_{[i]}\|^2.
\end{align}
and
\begin{align}
L_{\backslash [i]}(\vu_{\backslash [i+1]}^{*}) - L_{\backslash [i]}(\vu_{\backslash [i]}^{*}) &= |\va_{i+1}^{\T}\vu_{\backslash [i+1]}^{*}| - |\va_{i+1}^{\T}\vu_{\backslash [i]}^{*}| - \Delta_{[i]}^{\T}\va_{i+1}\sgli_{i+1}\nonumber\\
 &\hspace{5em}+ L_{\backslash [i+1]}(\vu_{\backslash [i+1]}^{*}) - L_{\backslash [i+1]}(\vu_{\backslash [i]}^{*})\nonumber\\
&\leq -\tfrac{1}{2} \|\Delta_{[i]}\|^2 + 2\|\Delta_{[i]}\|\cdot\|\va_{i+1}\|,
\label{eq:Lcavi_strongconv2}
\end{align}
where we use the fact $|\sgli_i|=1$ and Cauchy-Schwartz inequality in the last step.
From \eqref{eq:Lcavi_strongconv1} and \eqref{eq:Lcavi_strongconv2}, we can get $\|\Delta_{[i]}\|\leq2\|\va_{i+1}\|$.
Therefore, there exists $ c>0$, s.t., for any $\veps, D>0$,
\begin{align}
\P\left(|\va_{i}^{\T}\Delta_{[i]}|>\sqrt{\sampratio}\varepsilon\right) & \leq \P\left(|\va_{i}^{\T}\Delta_{[i]}|>\sqrt{\sampratio}\varepsilon\bigcap\|\Delta_{[i]}\|\leq D\right)+\P\left(\|\Delta_{[i]}\|>D\right)\nonumber\\
 & \leq\P\left(\left|\va_{i}^{\T}\tfrac{D\Delta_{[i]}}{\|\Delta_{[i]}\|}\right|>\sqrt{\sampratio}\veps\right)+\P\left(\|\va_{i+1}\|>\tfrac{D}{2}\right)\nonumber\\
 & \leq e^{-\tfrac{p{\sampratio}\veps^{2}}{2D^{2}}}+ce^{-c^{-1}p\left(\tfrac{D}{2\sqrt{\sampratio}}-1\right)_{+}^{2}},
 \label{eq:a1t_deltaUn1_concentrate2}
\end{align}
where $(x)_{+}\bydef \max\{x,0\}$. Then by choosing $D \asymp \sqrt{\sampratio}$ for small $\veps$ and $D \asymp \sqrt{\sampratio\veps}$ for large $\veps$, we can get (\ref{eq:a1t_deltaUn1_concentrate}).
\end{IEEEproof}
\begin{lem}
\label{lem:leave1_leavek_compare}There exists $c>0$, s.t., for any $b_{i}\in\R$,
$i=1,2,\ldots,k$ and $\veps>0$,
\begin{equation}
\P\left(\bigcap_{i=1}^{k}\left\{ \va_{i}^{\T}\vu_{\backslash i}^{*}\leq b_{i}\right\} \right)\geq\P\left(\bigcap_{i=1}^{k}\left\{ \va_{i}^{\T}\vu_{\backslash[k]}^{*}\leq b_{i}-\sqrt{\sampratio}\veps\right\} \right)-ck^{2}e^{-c^{-1}p\min\left\{ \tfrac{\veps^{2}}{k^{2}},\tfrac{\veps}{k}\right\} }\label{eq:leave1_leavek_compare_1}
\end{equation}
and
\begin{equation}
\P\left(\bigcap_{i=1}^{k}\left\{ \va_{i}^{\T}\vu_{\backslash i}^{*}\leq b_{i}\right\} \right)\leq\P\left(\bigcap_{i=1}^{k}\left\{ \va_{i}^{\T}\vu_{\backslash[k]}^{*}\leq b_{i}+\sqrt{\sampratio}\veps\right\} \right)+ck^{2}e^{-c^{-1}p\min\left\{ \tfrac{\veps^{2}}{k^{2}},\tfrac{\veps}{k}\right\} }.\label{eq:leave1_leavek_compare_2}
\end{equation}
\end{lem}
\begin{IEEEproof}
From Lemma \ref{lem:a1t_deltaUn1_concentrate}, for any $ k\in[p]$, there exists
$c>0$, s.t., for any $\veps>0$,
\begin{align*}
\P\left(\left|\va_{1}^{\T}(\vu_{\backslash1}^{*}-\vu_{\backslash[k]}^{*})\right| > \sqrt{\sampratio}\veps\right) & \leq\sum_{i=1}^{k-1}\P\left(\left|\va_{1}^{\T}(\vu_{\backslash[i]}^{*}-\vu_{\backslash[i+1]}^{*})\right|>\tfrac{\sqrt{\sampratio}\veps}{k-1}\right)\\
 & \leq cke^{-c^{-1}p\min\left\{ \tfrac{\veps^{2}}{k^{2}},\tfrac{\veps}{k}\right\} }.
\end{align*}
By the exchangeability of $\left\{ \va_{i}^{\T}\vu_{\backslash i}^{*},\va_{i}^{\T}\vu_{\backslash[k]}^{*}\right\} _{i\in[k]}$,
we have for any $ i\in[k]$, it holds that
\[
\P\left(\left|\va_{i}^{\T}(\vu_{\backslash i}^{*}-\vu_{\backslash[k]}^{*})\right| > \sqrt{\sampratio}\veps\right)\leq cke^{-c^{-1}p\min\left\{ \tfrac{\veps^{2}}{k^{2}},\tfrac{\veps}{k}\right\} }.
\]
Therefore, we have for any $ \veps>0$,
\begin{align*}
 \P\left(\bigcap_{i=1}^{k}\{\va_{i}^{\T}\vu_{\backslash i}^{*}\leq b_{i}\}\right)
&=  \P\left(\bigcap_{i=1}^{k}\{\va_{i}^{\T}\vu_{\backslash[k]}^{*}\leq b_{i}-\va_{i}^{\T}(\vu_{\backslash i}^{*}-\vu_{\backslash[k]}^{*})\}\right)\\
&\leq \P\left(\bigcap_{i=1}^{k}\{\va_{i}^{\T}\vu_{\backslash[k]}^{*}\leq b_{i}+\sqrt{\sampratio}\veps\}\right)+\P\left(\bigcup_{i=1}^{k}\{\left|\va_{i}^{\T}(\vu_{\backslash i}^{*}-\vu_{\backslash[k]}^{*})\right|>\sqrt{\sampratio}\veps\}\right)\\
&\leq \P\left(\bigcap_{i=1}^{k}\{\va_{i}^{\T}\vu_{\backslash[k]}^{*}\leq b_{i}+\sqrt{\sampratio}\veps\}\right)+ck^{2}e^{-c^{-1}p\min\left\{ \tfrac{\veps^{2}}{k^{2}},\tfrac{\veps}{k}\right\} },
\end{align*}
which is (\ref{eq:leave1_leavek_compare_1}). The other direction (\ref{eq:leave1_leavek_compare_2}) can be obtained in the same way.
\end{IEEEproof}
\subsubsection{$\big\{ \va_{i}^{\T}\vu_{\backslash[k]}^{*}\big\} _{i\in[k]} \tapprox{d} \big\{ \va_{i}^{\T}\widetilde{\vu}_{\backslash[k]}\big\} _{i\in[k]}$}
Next we show the joint distribution of $\big\{ \va_{i}^{\T}\vu_{\backslash[k]}^{*}\big\} _{i\in[k]}$
is close to $\big\{ \va_{i}^{\T}\widetilde{\vu}_{\backslash[k]}\big\} _{i\in[k]}$. First we show $\tfrac{\|\vu_{\backslash[k]}^{*}\|}{\sqrt{p}}\approx\tfrac{\|\widetilde{\vu}_{\backslash[k]}\|}{\sqrt{p}}=\optS$.
\begin{lem}
\label{lem:concentration_unotk_norm}When $k\leq\tfrac{p}{2}$, there exist $C,c>0$,
s.t. for any $\veps>0$,
\begin{equation}
\P\left(\left|\tfrac{\|\vu_{\backslash[k]}^{*}\|}{\sqrt{p}}-\optS\right|>\sqrt{\sampratio}\veps\right)
\leq \tfrac{c\sqrt{\sampratio}e^{-c^{-1}n\left(\veps-\tfrac{C k}{p}\right)_{+}^{2}}}{\max\left\{\veps-\tfrac{Ck}{p},n^{-\tfrac{1}{2}}\right\}}.
\label{eq:concentration_unotk_norm}
\end{equation}
\end{lem}
\begin{IEEEproof}
By the definition of $\vu_{\backslash[k]}^{*}$, we can get
\begin{align}
\tfrac{\|\vu_{\backslash[k]}^{*}\|}{\sqrt{p}} &= \tfrac{1}{\sqrt{p}} \min_{\vx\in[-1,1]^{p-k}}\|{\dmtx}_{\backslash[k]}\vx-({\dmtx}_{\backslash[k]}\vbeta_{\backslash[k]}+{\vw})\|\nonumber\\
&= \frac{\sampratio}{\delta_{p,\backslash[k]}}\cdot\frac{\min_{\vx\in[-1,1]^{p-k}}\|\widetilde{\dmtx}_{\backslash[k]}\vx-(\widetilde{\dmtx}_{\backslash[k]}\vbeta_{\backslash[k]}+\widetilde{\vw})\|}{\sqrt{p-k}},
\end{align}
where $\delta_{p,\backslash[k]}=\tfrac{n}{p-k}$, i.e., the
sampling ratio after removing $k$ predictors, $\widetilde{\dmtx}_{\backslash[k]}\iid\mathcal{N}\big(0,\tfrac{1}{p-k}\big)$
and $\widetilde{\vw}\sim\mathcal{N}\Big(\boldsymbol{0},\tfrac{\delta_{p,\backslash[k]} \nsdp^{2}}{\sampratio}\mI_{n}\Big)$. Define
\begin{align}
S_{p}^{*}(\delta) & \bydef\tfrac{\sampratio}{\delta}\min_{\tau>0}F_{p}\left(\tau;\tfrac{\delta\nsdp^{2}}{\sampratio},\delta\right),\label{eq:Spstar_alpha}
\end{align}
where $F_{p}$ is defined in (\ref{eq:F}). Similar to (\ref{eq:sqrt_Qpstar_convergence}), we
can get for $k\leq\tfrac{p}{2}$, $\exists c>0$, s.t., $\forall\veps>0$,
\begin{equation}
\P\left(\left|\tfrac{\|\vu_{\backslash[k]}^{*}\|}{\sqrt{p}}-S_{p,\backslash[k]}^{*}\right|>\veps\right)\leq \tfrac{c\sqrt{\sampratio}}{\veps}e^{-(p-k)\veps^{2}/c},\label{eq:equiv_notk_conv}
\end{equation}
where $S_{p,\backslash[k]}^{*}\bydef S_{p}^{*}(\delta_{p,\backslash[k]})$.

On the other hand, $|S_{p,\backslash[k]}^{*} - \optS|$ can be bounded
as follows. From (\ref{eq:Spstar_alpha}),
we can show when $k\leq\tfrac{p}{2}$, there exists $C>0$, s.t., $\left|\tfrac{dS_{p}^{*}(\delta)}{d\delta}\right|\leq \tfrac{C}{\sqrt{\sampratio}}$ for any $\delta\in [\sampratio,\delta_{p,\backslash[k]}]$.
Since $\optS = S_{p}^{*}(\sampratio)$ and $S_{p,\backslash[k]}^{*} = S_{p}^{*}(\delta_{p,\backslash[k]})$,
by the mean value theorem, we can get for $k\leq\tfrac{p}{2}$, there exists $C>0$, s.t.,
\begin{equation}
\left|S_{p,\backslash[k]}^{*} - \optS \right|\leq\tfrac{Ck\sqrt{\sampratio}}{p}.\label{eq:Sp_notk_Qpstar_diff}
\end{equation}
Now combining (\ref{eq:equiv_notk_conv})
, (\ref{eq:Sp_notk_Qpstar_diff}) and the condition $k\leq p/2$, we can obtain \eqref{eq:concentration_unotk_norm}.
\end{IEEEproof}
Based on Lemma \ref{lem:concentration_unotk_norm}, we can now show
$\va_{i}^{\T}\vu_{\backslash[k]}^{*}\approx \va_{i}^{\T}\widetilde{\vu}_{\backslash[k]}$,
if $k$ is not too large.
\begin{lem}
\label{lem:aiT_ustardiff_bd} If $k\leq\sqrt{p}$, then there exists $c>0$,
s.t., for any $\veps>0$ and $i\in[k]$,
\begin{equation}
\P\left(\left|\va_{i}^{\T}(\vu_{\backslash[k]}^{*}-\widetilde{\vu}_{\backslash[k]})\right|> \sqrt{\sampratio}\veps\right)\leq cp^{\tfrac{1}{2}}e^{-\sqrt{p}\veps/c}.\label{eq:aiT_ustardiff_bd}
\end{equation}
\end{lem}
\begin{IEEEproof}
Using \eqref{eq:concentration_unotk_norm} and following the similar steps as  \eqref{eq:a1t_deltaUn1_concentrate2}, we can get:
\begin{align}
\P\left(|\va_{i}^{\T}(\vu_{\backslash[k]}^{*}-\widetilde{\vu}_{\backslash[k]})|>\sqrt{\sampratio}\veps\right)  
\tleq{} Ce^{-\tfrac{p\veps^2}{2D^2}} + \tfrac{C\sqrt{\sampratio}e^{-C^{-1}n\left(\tfrac{D}{\sqrt{p}}-\tfrac{C k}{p}\right)_{+}^{2}}}{\max\left\{\tfrac{D}{\sqrt{p}}-\tfrac{Ck}{p},n^{-\tfrac{1}{2}}\right\}},\label{eq:aitDeltak_bd}
\end{align}
where $C$ is some constant. Setting $D = p^{\tfrac{1}{4}}\veps^{\tfrac{1}{2}}$ in \eqref{eq:aitDeltak_bd}, we can obtain \eqref{eq:aiT_ustardiff_bd}.
\end{IEEEproof}
Using Lemma \ref{lem:aiT_ustardiff_bd}, we can show that the joint distributions
of $\{ \va_{i}^{\T}\vu_{\backslash[k]}^{*}\} _{i\in[k]}$
and $\{ \va_{i}^{\T}\widetilde{\vu}_{\backslash[k]} \} _{i\in[k]}$ are similar.
\begin{lem}
\label{lem:leavek_leaveknormalized_compare}If $k\leq\sqrt{p}$, there exists $c>0$,
s.t., for any $b_{i}\in\R$, $i=1,2,\ldots,p$ and $\veps>0$,
\begin{align}
\P(\va_{i}^{\T}\vu_{\backslash[k]}^{*}\leq b_{i},i\in[k])\leq & \P\left(\va_{i}^{\T}\widetilde{\vu}_{\backslash[k]}\leq b_{i}+\sqrt{\sampratio}\veps,i\in[k]\right)+ckp^{\tfrac{1}{2}}e^{-\sqrt{p}\veps/c}\label{eq:leavek_leaveknormalized_compare_1}
\end{align}
and
\begin{align}
\P(\va_{i}^{\T}\vu_{\backslash[k]}^{*}\leq b_{i},i\in[k])\geq & \P\left(\va_{i}^{\T}\widetilde{\vu}_{\backslash[k]}\leq b_{i}-\sqrt{\sampratio}\veps,i\in[k]\right)-ckp^{\tfrac{1}{2}}e^{-\sqrt{p}\veps/c}.\label{eq:leavek_leaveknormalized_compare_2}
\end{align}
\end{lem}
\begin{IEEEproof}
The proof is similar to Lemma \ref{lem:leave1_leavek_compare} and is omitted here.
\end{IEEEproof}
\subsubsection{Proof of Proposition \ref{prop:approx_indep_Gaussian}}
The proof follows directly Lemma \ref{lem:leave1_leavek_compare} and Lemma \ref{lem:leavek_leaveknormalized_compare}.
\subsubsection{Proof of Proposition \ref{prop:joint_k_Bern_approx_indep_1}}
Letting $b_{i}=-A_{p}^{*}$ in (\ref{eq:joint_k_approx_indep_Gaussian}), we have
\begin{align}
\P\Big(\bigcap_{i=1}^{k}\left\{ \tilde{\sgli}_{i}\neq\beta_{i}\right\} \Big)
 & \geq \Phi^{k}\left(-\tfrac{1+\sqrt{\sampratio}\veps/A_{p}^{*}}{\opttau}\right)-{\Delta}_{p,k}\label{eq:joint_k_approx_indep_Gaussian_4_0}\\
 & \geq \Phi^{k}\left(-\tfrac{1}{\opttau}\right)\left[1-\tfrac{h({1}/{\opttau})\sqrt{\sampratio}\veps}{\opttau A_{p}^{*}}\right]^{k}-{\Delta}_{p,k},\label{eq:joint_k_approx_indep_Gaussian_4_1}
\end{align}
where $h(x)=\frac{\varphi(-x)}{\Phi(-x)}$ is the so-called inverse Mills ratio.  By (\ref{eq:Apstar_bd}), (\ref{eq:joint_k_approx_indep_Gaussian_4_1}) and (\ref{eq:inverse_mills_ratio_bd}) given in Appendix \ref{appen:mills}, there exists $c>0$, s.t., for any $ k\leq\sqrt{p}$ and small enough $\veps>0$,
\begin{align}
\P\Big(\bigcap_{i=1}^{k}\left\{ \tilde{\sgli}_{i}\neq\beta_{i}\right\} \Big) & \tgeq{\text{}}\Phi^{k}\left(-\tfrac{1}{\opttau}\right)\Big(1-\tfrac{ck\veps}{\sigma_{p}^{2}}\Big)-{\Delta}_{p,k}.\label{eq:joint_k_approx_indep_Gaussian_4}
\end{align}
On the other hand, we can also get the similar bounds as (\ref{eq:joint_k_approx_indep_Gaussian_4_0}) and (\ref{eq:joint_k_approx_indep_Gaussian_4}) for the other direction.

Now consider the case $k\leq p^{\tfrac{1}{8}}$. Accordingly, we set
$\veps=p^{-\tfrac{1}{4}}$. Then there exists $ c, c'>0$,
s.t.,
\begin{equation}
{\Delta}_{p,k}\leq c'p^{\tfrac{5}{8}}e^{-p^{1/4}/c'} \leq ce^{-p^{1/4}/c}.\label{eq:tilde_Delta_pk_order}
\end{equation}
As a result, from (\ref{eq:Apstar_bd}), (\ref{eq:joint_k_approx_indep_Gaussian_4_0}) and (\ref{eq:tilde_Delta_pk_order}), if $k\leq p^{\tfrac{1}{8}}$,
there exists $c>0$, s.t.,
\begin{equation}
\P\Big(\bigcap_{i=1}^{k}\left\{ \tilde{\sgli}_{i}\neq\beta_{i}\right\} \Big)\geq \Phi^{k}\Big(-\tfrac{1+cp^{-\tfrac{1}{4}}}{\opttau}\Big)-ce^{-p^{1/4}/c}.\label{eq:joint_k_Bern_approx_indep1}
\end{equation}
Meanwhile, we can also get for $\nsdp^2\geq \frac{c'}{\log^2 p}$,
\begin{align}
\P\Big(\bigcap_{i=1}^{k}\left\{ \tilde{\sgli}_{i}\neq\beta_{i}\right\} \Big)-\Phi^{k}\left(-\tfrac{1}{\opttau}\right) & \tgeq{\text{(a)}} -c\left[\Phi^{k}\left(-\tfrac{1}{\opttau}\right)kp^{-\tfrac{1}{4}}\polylog p + e^{-p^{1/4}/c}\right]\nonumber \\
& \tgeq{\text{(b)}} -c\Phi^{k}\left(-\tfrac{1}{\opttau}\right)\left[kp^{-\tfrac{1}{4}}\polylog p+\tfrac{e^{-p^{{1}/{4}}/c}}{\Phi^{k}\left(-\sqrt{\sampratio}/\nsdp\right)}\right]\nonumber\\
 & \tgeq{\text{(c)}} -\Phi^{k}\left(-\tfrac{1}{\opttau}\right)kp^{-\tfrac{1}{4}}\polylog p,\label{eq:joint_k_Bern_approx_indep_3}
\end{align}
where in step (a), we use \eqref{eq:joint_k_approx_indep_Gaussian_4}, in step (b), we use \eqref{eq:tau_t_ublb} and step (c) follows from inequality \eqref{eq:mills_ratio} and conditions $k\leq p^{\tfrac{1}{8}}$ and $\nsdp^2\geq \frac{c'}{\log^2 p}$. The other directions of (\ref{eq:joint_k_Bern_approx_indep1}) and (\ref{eq:joint_k_Bern_approx_indep_3}) can be derived similarly, which lead to (\ref{eq:joint_k_Bern_approx_indep_2}) and (\ref{eq:joint_k_Bern_approx_indep_1}). 
\subsection{Gaussian Tail Bounds}
\label{appen:mills}
Here we gather some properties of the Gaussian tail bounds that will be used in our proof. Let $\Phi(x)$ and $\varphi(x)$ be the CDF and PDF of the standard Gaussian distribution, respectively. It is well known that (see \cite[p.14]{vershynin2018high} for a proof), for any $x>0$,
\begin{equation}
\frac{1}{x}-\frac{1}{x^{3}} \leq m(x) \leq\frac{1}{x},\label{eq:mills_ratio}
\end{equation}
where $m(x) \bydef \frac{\Phi(-x)}{\varphi(-x)}$ is known as the Mills ratio. Correspondingly, the inverse Mills ratio is defined as $h(x)\bydef1/m(x)$. This provides us a way to approximate the tail probability $\Phi(-x)$ by $\varphi(x)$, which has an explicit form. In view of (\ref{eq:tau_t_ublb}) and (\ref{eq:mills_ratio}), there exists $M>1$,
s.t., for all $\eta\in[-1/2,1/2]$,
\begin{equation}
\tfrac{1+\eta}{\opttau}\leq h\Big(-\tfrac{1+\eta}{\opttau}\Big) \leq\tfrac{M(1+\eta)}{\opttau}.\label{eq:inverse_mills_ratio_bd}
\end{equation}
Meanwhile, from (\ref{eq:tau_t_ublb}) and (\ref{eq:inverse_mills_ratio_bd}), for all $\eta\in[-1/2,1/2]$,
\begin{align}
\Phi\left(-\tfrac{1+\eta}{\opttau}\right) & \leq\tfrac{1}{1+\eta}\sqrt{\tfrac{\nsdp^{2}}{\sampratio-1/2}}\tfrac{1}{\sqrt{2\pi}}{e^{-\frac{(1+\eta)^{2}(\sampratio-1/2)}{2\nsdp^{2}}}}\label{eq:Phi_1_tau_upbd}
\end{align}
and
\begin{align}
\Phi\left(-\tfrac{1+\eta}{\opttau}\right) & \geq\tfrac{1}{M(1+\eta)}\sqrt{\tfrac{\nsdp^{2}}{\sampratio}}\tfrac{1}{\sqrt{2\pi}}e^{-\frac{(1+\eta)^{2}(\sampratio-1/2+v_{p})}{2\nsdp^{2}}},\label{eq:Phi_1_tau_lowerbd}
\end{align}
where $v_{p} = \int_{b_p}^{\infty}x^{2}\Phi(dx)$, with $b_p = 2\sqrt{\tfrac{\sampratio-1/2}{\nsdp^2}}$. 
\subsection{An Auxiliary Result}
\label{subsec:pt_tildeNEB}
\begin{prop}
\label{prop:pt_tildeNEB}
As $p\to\infty$, it holds that
\begin{equation}
\lim_{p\to\infty}\P(\tneb=0)=\begin{cases}
1, & \liminf_{p\to\infty} \ptline>1,\\
0, & \limsup_{p\to\infty} \ptline<1.
\end{cases}\label{eq:tildeNe_transition}
\end{equation}
\end{prop}
\begin{IEEEproof}
When $\liminf_{p\to\infty}\ptline>1$,  ${\tfrac{4\nsdp^{2}}{\sampratio-1/2}}\leq \tfrac{2}{\log p}$ for large enough $p$. Combining (\ref{eq:joint_k_Bern_approx_indep_2}) and (\ref{eq:Phi_1_tau_upbd}) in Appendix \ref{appen:mills} gives us
\begin{align}
\E\tneb  & \leq Cp\Phi\left(-\tfrac{1+\eta}{\opttau}\right)+cpe^{-\sqrt[4]{p}/c}\nonumber \\
 & \leq \tfrac{C}{1+\eta}\sqrt{\tfrac{2}{\log p}} p^{1-\ptline[1+o(\eta)]},\label{eq:E_tildeNe_ubd}
\end{align}
where $\eta = -cp^{-\tfrac{1}{4}}$ and $C$ is some constant.  Therefore, from (\ref{eq:E_tildeNe_ubd}) and
Markov's inequality, $\lim_{p\to\infty}\P(\tneb\geq1)=0$.

When $\limsup_{p\to\infty} \ptline <1$, then
$\sigma_{p}\geq\frac{1}{\log p}$ for large enough $p$ and we have
\begin{align}
\E\tneb & \tgeq{\text{(a)}}p\Phi\left(-\tfrac{1}{\opttau}\right)\Big(1-p^{-\tfrac{1}{4}}\polylog p\Big)\nonumber \\
 & \tgeq{\text{(b)}} \tfrac{2\Big(1-p^{-\tfrac{1}{4}}\polylog p\Big)}{M\log p}e^{-\tfrac{v_p\log^2p}{2}}p^{1-\ptline},\label{eq:E_tildeNe_lbd}
\end{align}
where step (a) follows from (\ref{eq:joint_k_Bern_approx_indep_1}) and step (b) follows from (\ref{eq:Phi_1_tau_lowerbd}) in Appendix \ref{appen:mills}. In addition, it can be checked that $v_p$ as defined in (\ref{eq:tau_t_ublb}) satisfies $v_p\leq\tfrac{e^{-{b_p^{2}}/{2}}(b_p+2)^{2}}{2}$, where $b_p = {\tfrac{2\sqrt{\sampratio-1/2}}{\nsdp}}$. If  $\ptline\in\big[\tfrac{1}{2},1\big)$, $b_p \geq \tfrac{2\log p}{\sqrt{\sampratio-1/2}}$. Hence, there exists $C>0$, such that for large enough $p$, $v_p\leq \tfrac{1}{p^C}$. Then from (\ref{eq:E_tildeNe_lbd}) we can get $\lim_{p\to\infty} \E\tneb = \infty$. If $\ptline <\tfrac{1}{2}$, since $\opttau$ is strictly increasing with respect to $\nsdp^{2}$ as shown in (\ref{eq:tau_deri}), by step (a) above, it still
holds that $\lim_{p\to\infty} \E\tneb = \infty$.

We now prove that $\lim_{p\to\infty}\P(\tneb=0)=0$, when $\lim_{p\to\infty} \E\tneb = \infty$. The key lies in the approximate independence established in (\ref{eq:joint_k_Bern_approx_indep_1}).
First,
\begin{align}
\text{Var}\big(\tneb\big)= & \sum_{i=1}^{p}\text{Var}\left(\I_{\tilde{\sgli}_{i}\neq\sgli_{i}}\right)+\sum_{i\neq j}\text{Cov}\left(\I_{\tilde{\sgli}_{i}\neq\sgli_{i}},\I_{\tilde{\sgli}_{j}\neq\sgli_{j}}\right)\nonumber \\
\leq & \sum_{i=1}^{p}\P\left(\va_{i}^{\T}\vu_{\backslash i}^{*}\leq-A_{p}^{*}\right)\nonumber \\
 & +\sum_{i\neq j}\left|\P(\va_{i}^{\T}\vu_{\backslash i}^{*}\leq-A_{p}^{*},\va_{j}^{\T}\vu_{\backslash j}^{*}\leq-A_{p}^{*})-\P(\va_{i}^{\T}\vu_{\backslash i}^{*}\leq-A_{p}^{*})\P(\va_{j}^{\T}\vu_{\backslash j}^{*}\leq-A_{p}^{*})\right|\nonumber \\
\tleq{\text{(a)}} & \E\tneb\left(1+p^{-\tfrac{1}{4}}\polylog p\right)+(\E\tneb)^{2}p^{-\tfrac{1}{4}}\polylog p,\label{eq:var_tNe_upperbd1}
\end{align}
where in step (a), we have used (\ref{eq:joint_k_Bern_approx_indep_1}),
with $k=1,2$ and also (\ref{eq:E_tildeNe_ubd}). Let $\P(\tneb=0)=1-q_{p}$, $q_p\in [0,1]$. For any $p$, $\E\left(\tneb\mid\tneb>0\right)=\tfrac{\E\tneb}{q_{p}}$
and hence $\E\left(\tneb^{2}\mid\tneb>0\right)\geq\left(\tfrac{\E\tneb}{q_{p}}\right)^{2}$,
which indicates that $q_{p}\geq \tfrac{(\E\tneb)^{2}}{(\E\tneb)^{2}+\text{Var}(\tneb)}$.
This combined with \eqref{eq:var_tNe_upperbd1} and $\lim_{p\to\infty} \E\tneb = \infty$ leads
to: $\lim_{p\to\infty}q_{p}= 1$. Therefore, we conclude that $\lim_{p\to\infty}\P(\tneb=0)=0$.
\end{IEEEproof} 
\subsection{Proof of Proposition \ref{prop:Poisson_convergence}}
\label{appen:tneb_proof}

If $\ptline \geq2$, from (\ref{eq:E_tildeNe_ubd}) and (\ref{eq:Phi_1_tau_upbd}),
we know there exists $ c>0$, s.t., $\lambda_{p}\leq cp^{-\tfrac{1}{2}}$
and $\E\tneb\leq cp^{-\tfrac{1}{2}}$. Hence, $d_{\text{TV}}(\tneb,\mathscr{P}_{\optlam})$ can be bounded
as:
\begin{align*}
 d_{\text{TV}}(\tneb,\mathscr{P}_{\optlam}) & \leq\frac{1}{2}\left|\P\left(\tneb=0\right)-e^{-\optlam}\right|+\frac{1}{2}\P\left(\tneb\geq1\right)+\frac{1}{2}\left(1-e^{-\optlam}\right) \leq 2cp^{-\tfrac{1}{2}}.
\end{align*}
On the other hand, if $\ptline<2$,
then for large enough $p$, it holds that $\sigma_{p}\geq\frac{1}{\log p}$.
Choose $L$ in (\ref{eq:Bonferroni_1}) to be $L=\left\lfloor 5\log p\right\rfloor .$
Without loss of generality, assume $L-k$ is odd (otherwise we add
$L$ by $1$). Then from Bonferroni's inequality (\ref{eq:Bonferroni_1}), for $k\leq\left\lfloor \log p\right\rfloor $,
\begin{align}
\P\left(\tneb=k\right) & \leq\sum_{m=0}^{L-1-k}\tfrac{(-1)^{m}}{k!m!}p^{m+k}S_{[m+k]}\nonumber \\
 & \tleq{\text{(a)}}\sum_{m=0}^{L-1-k}\tfrac{(-1)^{m}}{k!m!}p^{m+k}\Phi^{m+k}\left(-\tfrac{1}{\opttau}\right)+\sum_{m=0}^{L-1-k}\tfrac{p^{m+k}\Phi^{m+k}\left(-\tfrac{1}{\opttau}\right)}{k!m!}p^{-1/4}L_{p}\nonumber \\
 & \tleq{\text{(b)}}\tfrac{\lambda_{p}^{k}}{k!}e^{-\lambda_{p}}\left(1+\left(\tfrac{\lambda_{p}e}{L-k}\right)^{L-k}e^{\lambda_{p}-1}\right)+\tfrac{\lambda_{p}^{k}}{k!}e^{\lambda_{p}}p^{-1/4}L_{p}\nonumber \\
 & \tleq{\text{(c)}}\frac{\lambda_{p}^{k}}{k!}e^{-\lambda_{p}}\left[1+\left(\tfrac{C}{{\log^{2} p}}\right)^{\log p}+p^{-1/5}L_{p}\right].\label{eq:P_k_upper_bd}
\end{align}
Here, $L_p$ is the shorthand notation for a term of order $\mathcal{O}(\polylog p)$ and $C$ is some constant, step (a) follows from (\ref{eq:joint_k_Bern_approx_indep_1}),
in step (b) we use Taylor approximation and inequality
$n!\geq e\left(\tfrac{n}{e}\right)^{n}$ and step (c) follows from conditions $L=\left\lfloor 5\log p\right\rfloor$, $k\leq\left\lfloor \log p\right\rfloor$ and $\limsup_{p\to\infty}\tfrac{\optlam}{\sqrt{\log p}} < \infty$. In a similar manner, for the other direction, we can also obtain
\begin{align}
\P\left(\tneb=k\right) & \geq\frac{\lambda_{p}^{k}}{k!}e^{-\lambda_{p}}\left[1-\left(\tfrac{C}{{\log^{2} p}}\right)^{\log p}-p^{-1/5}L_{p}\right].\label{eq:P_k_lower_bd}
\end{align}
By (\ref{eq:P_k_upper_bd}) and (\ref{eq:P_k_lower_bd}), for $k\leq \left\lfloor \log p\right\rfloor$,
\begin{equation}
\left|\P\left(\tneb=k\right)-\tfrac{\lambda_{p}^{k}}{k!}e^{-\lambda_{p}}\right|\leq\tfrac{\lambda_{p}^{k}}{k!}e^{-\lambda_{p}}p^{-1/5}L_{p}.\label{eq:P_k_bd_1}
\end{equation}
Then $d_{\text{TV}}(\tneb,\mathscr{P}({\lambda_{p}}))$
can be bounded as:
\begin{align*}
d_{\text{TV}}(\tneb,\mathscr{P}({\lambda_{p}})) & \leq\tfrac{1}{2}\sum_{k=0}^{\lfloor\log p\rfloor}\left|\P(\tneb=k)-\tfrac{\lambda_{p}^{k}e^{-\lambda_{p}}}{k!}\right|+\tfrac{1}{2}\sum_{k=\lfloor\log p\rfloor+1}^{\infty}\P(\tneb=k)+\sum_{k=\lfloor\log p\rfloor+1}^{\infty}\tfrac{\lambda_{p}^{k}e^{-\lambda_{p}}}{2k!}\\
 & \tleq{\text{(a)}}\tfrac{p^{-1/5}L_{p}}{2}\sum_{k=0}^{\lfloor\log p\rfloor}\tfrac{\lambda_{p}^{k}e^{-\lambda_{p}}}{k!}+\tfrac{1}{2}\left[1-\sum_{k=0}^{\lfloor\log p\rfloor}\tfrac{\lambda_{p}^{k}e^{-\lambda_{p}}}{k!}(1-p^{-\tfrac{1}{5}}L_{p})\right]+\sum_{k=\lfloor\log p\rfloor+1}^{\infty}\tfrac{\lambda_{p}^{k}e^{-\lambda_{p}}}{2k!}\\
 & \tleq{\text{(b)}}p^{-1/5}L_{p},
\end{align*}
where in step (a) we use (\ref{eq:P_k_bd_1}), in step (b) we use
Chernoff's bound for the tail probability of Poisson random variables \cite[p.20]{vershynin2018high}: for $X\sim \mathscr{P}(\lambda), k>\lambda$, $\P(X>k) \leq e^{-\lambda}\left(\frac{e\lambda}{k}\right)^{k}$ and the condition that $\limsup_{p\to\infty}\tfrac{\optlam}{\sqrt{\log p}} < \infty$.

\bibliographystyle{IEEEtran}
\bibliography{refs}

\end{document}